\documentclass[sigconf]{acmart}
\usepackage[utf8]{inputenc}
\usepackage{graphicx}

\usepackage{dutchcal}
\usepackage[mathscr]{eucal}
\usepackage{latexsym}

\usepackage{setspace}
\usepackage{commath}
\usepackage[english]{babel}
\usepackage[noend]{algpseudocode}
\usepackage{algorithmicx}
\usepackage{booktabs}
\usepackage{algorithm}
\usepackage{dsfont}
\usepackage[utf8]{inputenc}
\usepackage{mathtools}
\usepackage{centernot}
\usepackage{makecell}
\usepackage{tabularx}
\usepackage{xcolor}
\usepackage{mathtools}
\usepackage{soul}
\usepackage{hyperref}

\newcommand{\name}{VDR}

\newcommand\vldbdoi{XX.XX/XXX.XX}

\newcommand\vldbavailabilityurl{URL_TO_YOUR_ARTIFACTS}

\begin{document}
\title{A Neural Approach to Spatio-Temporal Data Release with User-Level Differential Privacy}

\author{Ritesh Ahuja}
\thanks{*Equal contribution.}
\affiliation{\institution{University of Southern California\parbox{2pt}{\vspace*{-1cm}\hspace*{-1.3cm}*}}}
\email{riteshah@usc.edu}
\author{Sepanta Zeighami}
\affiliation{\institution{University of Southern California\parbox{2pt}{\vspace*{-1cm}\hspace*{-0.8cm}*}}}
\email{zeighami@usc.edu}
\author{Gabriel Ghinita}
\affiliation{\institution{Hamad Bin Khalifa University}}
\email{gghinita@hbku.edu.qa}
\author{Cyrus Shahabi}
\affiliation{\institution{University of Southern California}}
\email{shahabi@usc.edu}
\begin{abstract}
Several companies (e.g., Meta, Google) have initiated ``data-for-good'' projects where aggregate location data are first sanitized and released publicly, which is useful to many applications in transportation, public health (e.g., COVID-19 spread) and urban planning. {\em Differential privacy (DP)} is the protection model of choice to ensure the privacy of the indivduals who generated the raw location data. 
However, current solutions fail to preserve data utility when each individual contributes multiple location reports (i.e., under user-level privacy). To offset this limitation, public releases by Meta and Google use high privacy budgets (e.g., $\varepsilon=10$-$100$), resulting in poor privacy. We propose a novel approach to release spatio-temporal data privately and accurately. 
We employ the pattern recognition power of neural networks, specifically variational auto-encoders (VAE), to reduce the noise introduced by DP mechanisms such that accuracy is increased, while the privacy requirement is still satisfied. Our extensive experimental evaluation on real datasets shows the clear superiority of our approach compared to benchmarks.
\end{abstract}

\newcommand{\revision}[1]{\textcolor{black}{#1}}

\maketitle

\pagestyle{plain}



\section{Introduction}\label{sec:intro}

Several ``data-for-good'' projects \cite{aktay2020google, dataforgood, safegraph_patterns} initiated by major companies (e.g., Meta, Google) release to the public spatio-temporal datasets to benefit COVID-19 spread modeling~\cite{chang2021mobility,rambhatla2022toward, zeighami2021estimating} and understand human mobility \cite{bassolas2019hierarchical, fatehkia2020relative}. Most often, spatio-temporal data are provided in the form of snapshot high resolution population density information, where the released statistics capture population counts in small areas for short time periods. Since high resolution is required for utility (e.g., in modeling COVID hotspots) privacy risks are elevated. To prevent malicious actors from using the data to infer sensitive details about individuals, the released datasets must be first sanitized. Typically, ~\cite{aktay2020google, censusgovKDD18, dataforgood, safegraph_patterns}, {\em differential privacy (DP)} is employed as protection model, due to its formal  protection guarantees that prevent an adversary to learn whether a particular individual's data has been included in the release or not.

Most existing work on DP-compliant publication of location data focused on \textit{single snapshot} releases, where each individual contributes a {\em single location report} ({\em event-level} privacy)~\cite{zhang2017privbayes, xiao2012dpcube, mckenna2018optimizing, hay2009boosting, qardaji2013differentially, cormode2012differentially, zhang2016privtree, li2014data, acs2012differentially, qardaji2013understanding}. When releasing \textit{multiple} snapshots, the ability of an adversary to breach privacy increases significantly, and a shift to {\em user-level} privacy~\cite{dwork2008differential} is required. To protect privacy in this scenario, an increased amount of noise is needed, which often grows linearly in the number of user contributions. Only a handful of techniques~\cite{acs2014case,hien-tmc} considered spatio-temporal location data releases, and their accuracy is subpar. 
Existing industry projects use simple DP mechanisms that 
do not account for specific dataset characteristics \cite{aktay2020google, dataforgood, safegraph_patterns, bassolas2019hierarchical, houssiau2022difficulty}. The amount of privacy budget spent for such data releases is often not reported, or it is excessive~\cite{bassolas2019hierarchical, houssiau2022difficulty, censusgovKDD18}, thus providing insufficient protection.
Reports of incorrect privacy accounting in such releases~\cite{bassolas2019hierarchical, houssiau2022difficulty} 
further necessitate a thorough end-to-end study of custom DP algorithms for spatio-temporal data.

Two key aspects must be addressed. First, one needs to bound {\em sensitivity} (see Section 2 for a formal definition) by limiting the number of location reports from any single user, which can be achieved through sampling.  Density information must be adjusted to account for the fact that it is calculated on a subset of the actual data. Second, the effect of noise added by DP mechanisms must be addressed. Such mechanisms consider the worst-case scenario over all possible data distributions and query configurations, and err on the safe side, adding more noise than strictly necessary. A {\em denoising} post-processing step that leverages spatio-temporal data characteristics can significantly boost accuracy, while still satisfying privacy. Recent advances in neural networks, such as variational auto-encoders (VAE), are good at capturing complex density patterns, and can enable effective denoising.

We propose {\em VAE-based Density Release} (\name{}), a system specifically designed for accurate, DP-compliant release of spatio-temporal data.
Noisy spatio-temporal data histograms exhibit patterns akin to visual patterns in image sequences. This observation allows one to leverage a vast amount of work on image pattern recognition and apply it to spatio-temporal data. \name{} sanitizes density information by adding DP-compliant noise, then improves  accuracy by performing a post-processing {\em denoising} step based on convolutional neural networks (CNN). We utilize {\em variational auto-encoders (VAE)} to capture data patterns without fitting to the noise. We employ multi-resolution learning to capture patterns at multiple granularities, improving accuracy for a broad range of query extents.

To reduce sensitivity of user-level privacy, we reduce the number of input samples from any individual in a DP-compliant way. To counter-balance the effect of sampling, we design a novel private statistical estimator which scales up query results to preserve accuracy. This permits us to control the sensitivity in user-level privacy without significantly affecting accuracy.

Our specific contributions are:
\begin{itemize}
    \item We propose an end-to-end privacy-preserving system for spatio-temporal datasets that satisfies user-level differential privacy and preserves data accuracy;
    \item We introduce a novel approach to user-level sampling that reduces sensitivity while preserving density information across time;
    \item We design a novel denoising technique that uses variational auto-encoders and image feature extraction concepts to accurately model patterns in spatio-temporal data;
    \item We design a technique to offset the effects of location sampling in order to provide accurate query answers; to that extent, we employ DP-compliant statistical estimators;
    \item We perform an extensive experimental evaluation on real data which shows that the proposed approach clearly outperforms all existing approaches.
\end{itemize}

 We provide background information and formulate the studied problem in Section 2. Section 3 introduces the proposed sampling and denoising techniques. Section 4 explores system design trade-offs. Section 5 presents the experimental evaluation results. We survey related work in Section 6 and conclude in Section 7.

\section{Preliminaries}\label{sec:bck}
\subsection {Differential Privacy}
{Given {\em privacy budget} $\varepsilon \in (0, +\infty)$}, a mechanism $\mathscr{M}$ satisfies $\varepsilon$-differential privacy~\cite{dwork2014algorithmic} iff for any {\em sibling} datasets $D$ and $D^\prime$  differing in a single tuple, and for all $E \subseteq$ Range($\mathscr{M}$)
\begin{equation}
\vspace*{-4pt}
\text{Pr}[\mathscr{M}(D) \in E] \leq e^\varepsilon \text{Pr}[\mathscr{M}(D^\prime) \in E]
\end{equation}

The protection provided by DP is stronger when $\varepsilon$ approaches $0$. 
The {\em sensitivity} of a function (e.g., a query) $f$, denoted by $\Delta_f$, is the maximum amount the value of $f$ can change when adding or removing a single individual's contribution from the data. The $\varepsilon$-DP guarantee can be achieved by adding random noise derived from the Laplace distribution $\text{Lap}(\Delta_f/\varepsilon)$. For a query $f:D \rightarrow \mathbb{R}$, the {\em Laplace mechanism} (LPM) returns $f(D) + \text{Lap}(\Delta_f/\varepsilon)$, where Lap$(\Delta_f/\varepsilon)$ is a sample drawn from the probability density function Lap$(x|(\Delta_f/\varepsilon)) = (\varepsilon/2\Delta_f) \text{exp}(-|x|\varepsilon/\Delta_f)$ \cite{dwork2014algorithmic}. 

Let $f$ be a vector-valued function that outputs the population count in a location histogram at {\em every} time snapshot. With user-level privacy, removing an individual's data may cause changes in multiple elements of $f$ (in the worst case, the maximum number of reports across all individuals). Contrast this with event-level privacy, where sibling datasets differ in a single value. User-level privacy causes a significant increase in sensitivity, which must be carefully controlled to prevent utility loss.



\subsection{Problem Formulation}
We are given a dataset $D$ consisting of user location reports with four attributes: latitude ({\em lat}), longitude ({\em lon}), timestamp ({\em time}) and user id. The goal is to release high-resolution density information of $D$ for arbitrary spatial regions over time. We build a  $\mathbb{M}$$\times$$\mathbb{M}$$\times$$\mathbb{T}$ histogram $H$ over the data, where $\mathbb{M}$ and $\mathbb{T}$ determine the spatial and temporal resolution. $M$ and $\mathbb{T}$ are determined by application requirements, e.g., release a histogram at 30$\times$30m resolution and one hour time granularity over a duration of 24 hours~\cite{dataforgood, li2014data}. We design a mechanism $\mathscr{M}$ that takes $H$ as an input and outputs a histogram $\hat{H}$, where $\mathscr{M}$ preserves $\varepsilon$-DP.  We focus on the following three  statistical query types:

\textbf{Range count queries}. Given a query range, defined by minimum and maximum values (i.e., a range) for $lat$, $lon$ and $time$, find the number of user location reports in $D$ that satisfy this range predicate. For a query $q$, we measure the utility of its estimated DP-preserving answer, $y$, compared to the true answer $u$ using the \textit{relative error metric}, defined as $\text{RE}(y, u)=\frac{|y-u|}{\max\{u, \psi\}}$, where $\psi$ is a smoothing factor necessary to avoid division by zero.

\textbf{Nearest hot-spot queries}. Given a query location $q$ (lat, lon, time), a density threshold, $\nu$, and a spatio-temporal extent $SR$ (defined by a time duration and lengths of lat and lon geo-coordinates), find the closest cell to query $q$ within extents $SR$ that contains a number of at least $\nu$ locations signals. The hotspot query may be answered using an expanding search in the 3-d histogram until a cell within $SR$ having at least $\nu$ points is found. If none is found, the cell with the maximum count is reported. We evaluate this query in two ways: the distance penalty is measured as the Mean Absolute Error (MAE) between the true distance (as computed on $H$) and reported distance (computed on $\hat{H}$) to the hotspot. To capture hotspot density estimation errors, we measure \textit{Regret}, defined as the deviation of the reported density of the found hotspot (on noisy histogram $\hat{H}$) from the specified threshold $\nu$. Regret for a query is zero if the reported hotspot meets the density threshold. 

\textbf{Forecasting query}. Given a timeseries of density counts for a 2-d region (defined with minimum and maximum $lat$ and $lon$ values), and a forecasting horizon $h$ {\em not covered} within the timeseries, predict the count of location reports for $h$ future timesteps. To evaluate this query, we utilize \textit{holdout testing}, which removes the last $h$ data points of the timeseries, calculates the forecasting model fit for the remaining historical data, makes forecasts for $h$ timesteps, and compares the error between the forecast points and their corresponding, \textit{held-out}, data points. We report the symmetric mean absolute percentage errors (sMAPE) as $\text{sMAPE} = \frac{1}{h}\sum_{t=1}^{h}\frac{\left|F_t-A_t\right|}{(A_t+F_t)/2}$, where $A_t$ are the true counts from $H$ in the $h$ timesteps and $F_t$ are the $h$ predicted counts from a forecasting algorithm fitted to the historical data points from $\hat{H}$. 





\subsection{Data Characteristics and Assumptions}\label{sec:data_model}


\revision{{\bf Density patterns.} 
Real-world location datasets are generated from human mobility and tend to have similar density patterns across cities \cite{yang2015nationtelescope}, exhibiting homogeneous characteristics that capture activities like travel on road networks, visits to bars and restaurants, etc.
Such density patterns exist across space. For instance, areas with high density (bars, restaurant, malls) may be separated by sparse residential areas, and connected by road networks. The density patterns may also exist over time. For instance, an area that is busy on a weekend night may be less populated during a weekday afternoon (we quantify existence of temporal patterns in our datasets in Sec.~\ref{subsec:settings}). 
VDR exploits the spatial and/or temporal patterns in such datasets to achieve high accuracy 
compared with existing work on a broad array of real datasets, as we show in Sec.~\ref{sec:exp}. 
}

\noindent
\revision{\textbf{Data Distribution across users.} In real-world location datasets, the number of contributed data points varies significantly across users. While most users contribute few points, some prolific users may contribute a very large number of points. The number of points a user contributes often follows a power law \cite{muchnik2013origins, yang2015nationtelescope}. We observed this power law across all our datasets used in our experiments (see Sec.~\ref{subsec:settings}). For example, for a location dataset in Houston, the maximum number of points contributed by a user is $90,676$, while 80\%  of users have at most 100 points. This is because, in real datasets, location updates are often collected from mobile apps, with the amount of user contributions varying due to different app utilization across users. Such a power law distribution is taken into account in our statistical refinement (Sec.~\ref{sub:refinement}) to improve accuracy while accounting for user-level privacy.}

\section{VAE-based Density Release (VDR)}\label{sec:stdata}

\if {0}{
Our approach consists of three steps: (1) {\em data collection},
(2) {\em learned denoising} 
and (3) {\em statistical refinement}.
We provide technical descriptions of each step in Sections~\ref{sub:collection}~-~\ref{sub:refinement}. 
}\fi

\begin{figure}
    \centering
    \includegraphics[width=\columnwidth]{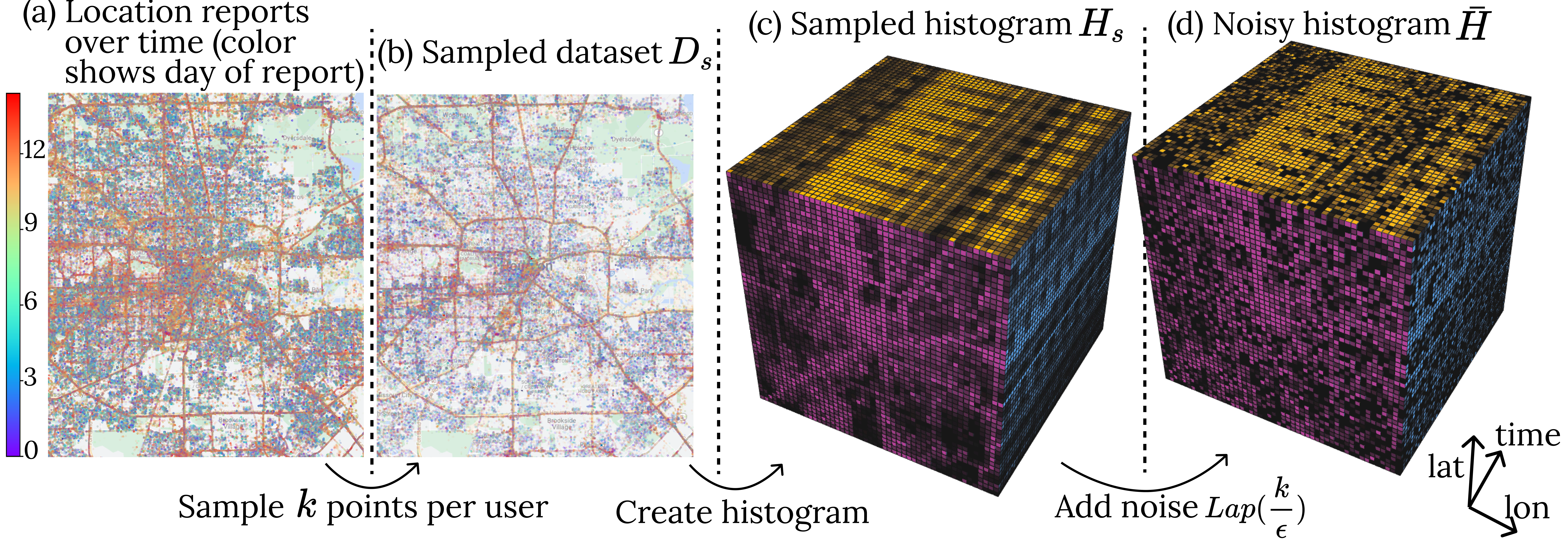}
    \caption{(a) and (b): real-world complete and sampled dataset of location reports over time in Houston. (c) and (d): exact and noisy 3-d histograms created from the sampled dataset, higher brightness shows higher density.}
    \label{fig:data_collection}
\end{figure}

\subsection{Data Collection}\label{sub:collection}
Data collection uses a combination of sampling and noise addition to create a differentially private histogram of the data without making any modelling assumptions. In the case of spatio-temporal data, simple noise addition will lead to poor quality results, as the amount of noise needed will destroy any meaningful signal in the data. We first discuss the naive solution and its specific challenge for spatiotemporal data; subsequently, we show how sampling is able to improve the accuracy; finally we present the details of the data collection mechanism. We use as running example a real world dataset of location reports from Houston, TX, USA (see Section \ref{sec:exp} for exact details of dataset). 

\textbf{DP Histogram Release}. Given dataset $D$  with location reports from different users over time, the goal is to create a histogram of the data while preserving $\varepsilon$-DP. One way to do this (without making any modeling assumptions) is to first create the true histogram of the data $H$, and then add independent Laplace noise, $Lap(\frac{\Delta}{\varepsilon})$ to each cell of the histogram, where $\Delta$ is the sensitivity of the query of number of data points falling inside a cell. This sensitivity is equal to the maximum number of points, $k_{\max}$, a user contributes to the dataset. Thus, the DP histogram of the data can be written as $\bar{H}=H+Lap(\frac{k_{\max}}{\varepsilon})$, where independently generated random noise is added to each cell of the histogram.

\textbf{Challenge for Spatiotemporal Histograms}. 
\revision{As discussed in Sec.~\ref{sec:data_model}, real-world location datasets follow power law distribution. For the Houston dataset, $k_{max}=90,676$, while 80\%  of users have at most 100 points. In this dataset, for ranges of 30 meters and 1 hour time periods, only 1 percent of the histogram cells have values more than 25. In such datasets, applying the above DP histogram method, without adjusting for the power law distribution across users, leads to poor accuracy. In our running example,} Laplace noise scaled to $k_{\max}/\varepsilon$, for any reasonable value of $\varepsilon$, wipes out any meaningful information in all but a few cells.

\textbf{Sampling to Bound Sensitivity}. Instead of using all the data points of users when creating the histogram, we sample a maximum of ${k}$ points per user, for a user parameter ${k} < k_{\max}$. Specifically, we sample a subset of points $D_s \subseteq D$ as follows: for any user with more than ${k}$ points, we sample ${k}$ of their points uniformly at random. For users with at most ${k}$ points, we keep all their points. This reduces the sensitivity of releasing histogram to ${k}$, requiring that we add only noise $Lap(\frac{{k}}{\epsilon})$ to each cell of the histogram. In this way, we can exploit the skewness in user contributions to the dataset, because by setting ${k}$ to a small value, we are able to retain most of the original data. For the Houston dataset, setting ${k}$ to 128 captures nearly 25\% percent of the data, while reducing sensitivity by 700\%. Consequently, we bias the data distribution in order to reduce variance in query reporting. Nonetheless, sampling introduces \textit{sampling error} in the answers over  histogram  $D_s$. In our statistical refinement step (Sec.~\ref{sub:refinement}), we discuss how we can counter this source of error. We further discuss trade-offs arising in our method based on the choice of $k$ in Sec.~\ref{subsec:system}.




\textbf{Summary and Example}. Data collection is depicted in Fig.~\ref{fig:data_collection} for our running example, where we release a noisy $50$$\times$$50$$\times50$ histogram of the dataset of location reports in Houston (\revision{granularity chosen for better visualization}). We sample up to ${k}$ for each user from the complete dataset $D$ to create the sampled dataset $D_s$. Then, we create a 3-dimensional histogram, $H_s$ of $D_s$. Finally, we create the histogram $\bar{H}=H_s+Lap(\frac{{k}}{\varepsilon})$ so that the data collection process satisfies $\varepsilon$-DP. The output of the data collection step is the noisy histogram $\bar{H}$. The output in Fig.~\ref{fig:data_collection} (d) shows the noisy histogram, where each cell corresponds to noisy density in a 800$\times$800 meter cell for a 4 hour time period.

\begin{figure}
    \centering
    \includegraphics[width=\columnwidth]{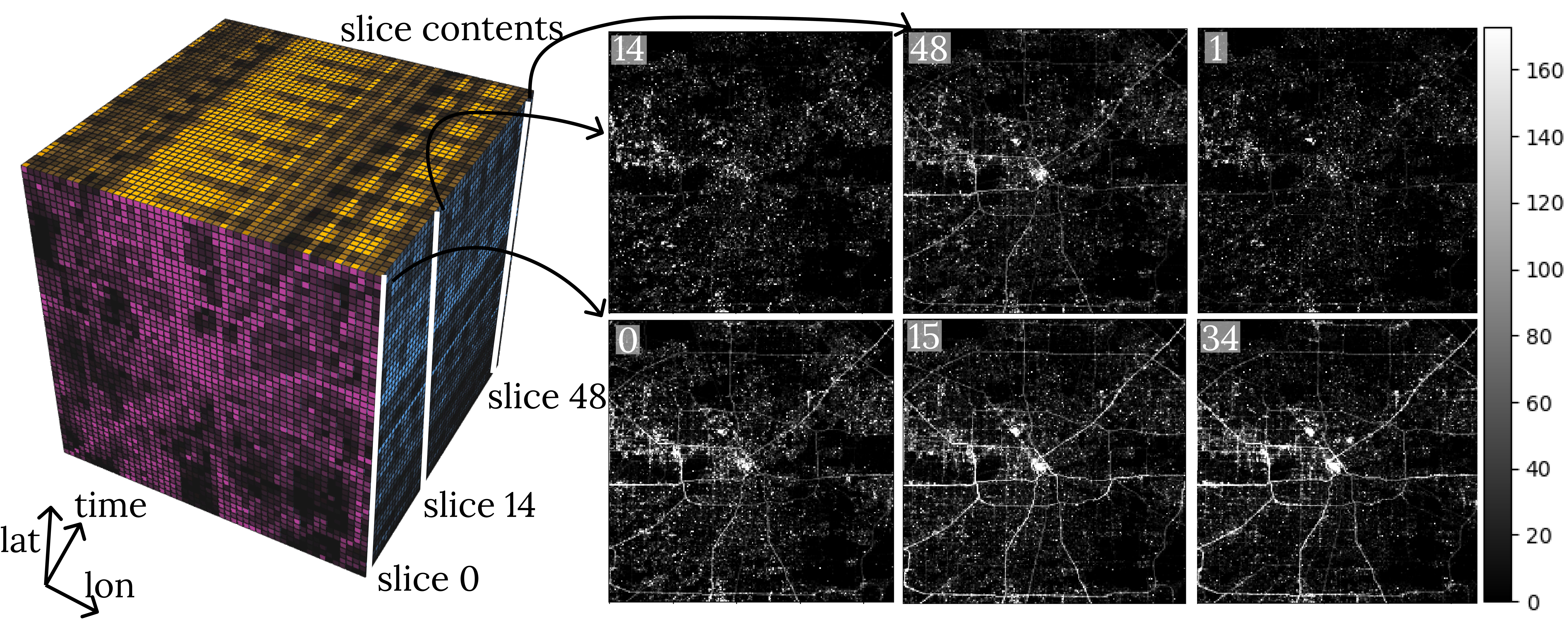}
    \caption{Spatial patterns over time on histogram slices}
    \label{fig:histogram_overtime}
\end{figure}

\subsection{Learned Denoising}\label{sub:denoising}
Denoising uses machine learning, specifically VAEs, to identify spatial-temporal patterns, and  utilizes them to improve histogram accuracy. Our main observations are: (1) spatio-temporal histograms are similar in nature to a sequence of images, thus methods from image representation learning can be applied to capture data patterns; (2) regularized representation learning can ensure the model learns a denoised representation of the data while not over-fitting noise; and (3) multi-resolution learning can capture spatio-temporal patterns at different granularities. 

\subsubsection{Design Principles\nopunct}\hfill\\

\textbf{Denoising with Regularized Representation Learning}. We want to derive a denoised histogram $\hat{H}$ from $\bar{H}$ that is similar to $H$, where similarity is measured as norm $\lVert H-\hat{H}\rVert$, i.e., the sum of squared differences across all cells of the histograms. To achieve this, consider a function {\tt encoder}($\bar{H}$) that creates an encoding, $z$, of the noisy histogram, and a function {\tt decoder}($z$), that outputs a histogram, $\hat{H}$, from the encoding $z$. Consider the problem of learning an encoding $z$ (i.e., by learning functions {\tt encoder}(.) and {\tt decoder}(.)), so that $\lVert\hat{H}-\bar{H}\rVert$ is minimized, where we call $\lVert\hat{H}-\bar{H}\rVert$ the {\em reconstruction error}.
Our goal is to obtain an encoding $z$ that summarizes the patterns in $\bar{H}$, since such patterns will also exist in the true histogram $H$. 
To see why this is possible, observe that a constraint on $z$ limits its representation power. For instance, by setting the dimensionality of $z$ to be lower than that of $\bar{H}$, $z$ cannot contain as much information as $\bar{H}$. Thus, a regularized encoding, $z$, that minimizes the reconstruction error cannot contain all the information in $\bar{H}$. By learning a regularized representation, $z$, a model is able to capture the patterns in $\bar{H}$ that best summarize the histogram. Such a summary \revision{will be less noisy}, as \revision{noise} does not generalize across the histogram (noise is independently added to each cell) and can only be memorized individually per cell. Thus, by properly regularizing the encoding, we can find an encoding that is denoised, i.e., contains the patterns in the data, but \revision{less} noise. Subsequently, by decoding such a representation, we can obtain a denoised histogram. That is, the regularization ensures that even though we try to minimize the reconstruction error $\lVert\hat{H}-\bar{H}\rVert$, we obtain a histogram such that $\lVert\hat{H}-{H}\rVert$ is smaller than $\lVert\bar{H}-{H}\rVert$.

\textbf{Spatial Patterns as Visual Patterns}. Denoising with regularized representation learning will be beneficial only if the model is able to extract the patterns in the histogram. To facilitate this, we observe that a 3-d histogram of the data can be seen as a sequence of images, as shown in Fig.~\ref{fig:histogram_overtime} for our running example. The left side of Fig.~\ref{fig:histogram_overtime} plots the 3-d histogram which represents a time-series of two-dimensional histograms, one per each timestamp. We call each of these 2-d histograms a \textit{slice}. On the right side of the figure, we plotted various slices, each corresponding to a different timestamp. We can see that spatial patterns in the histogram are in fact visual patterns. For instance, patterns corresponding to roads or busy areas can be seen as lines or blobs in the image. \revision{In real world location datasets, we expect such consistent patterns across space that may also be repeating over time (see Sec.~\ref{sec:data_model} regarding location data characteristics)}, suggesting that representation learning can be achieved effectively using techniques from image feature learning. 

\textbf{Multi-Resolution Learning (MRL) at Varying Granularity}. Spatial patterns in the data exist at various granularities of the input histogram. 
Patterns at finer resolutions feature roads more prominently, while patterns at coarser granularities feature primarily neighbourhood densities. Furthermore, the patterns in coarser granularity histograms are less affected by noise, which allows the model to still infer spatial density. We propose to train a single model based on data configured at multiple granularities to improve denoising accuracy. 

\begin{figure}
    \centering
    \includegraphics[width=\columnwidth]{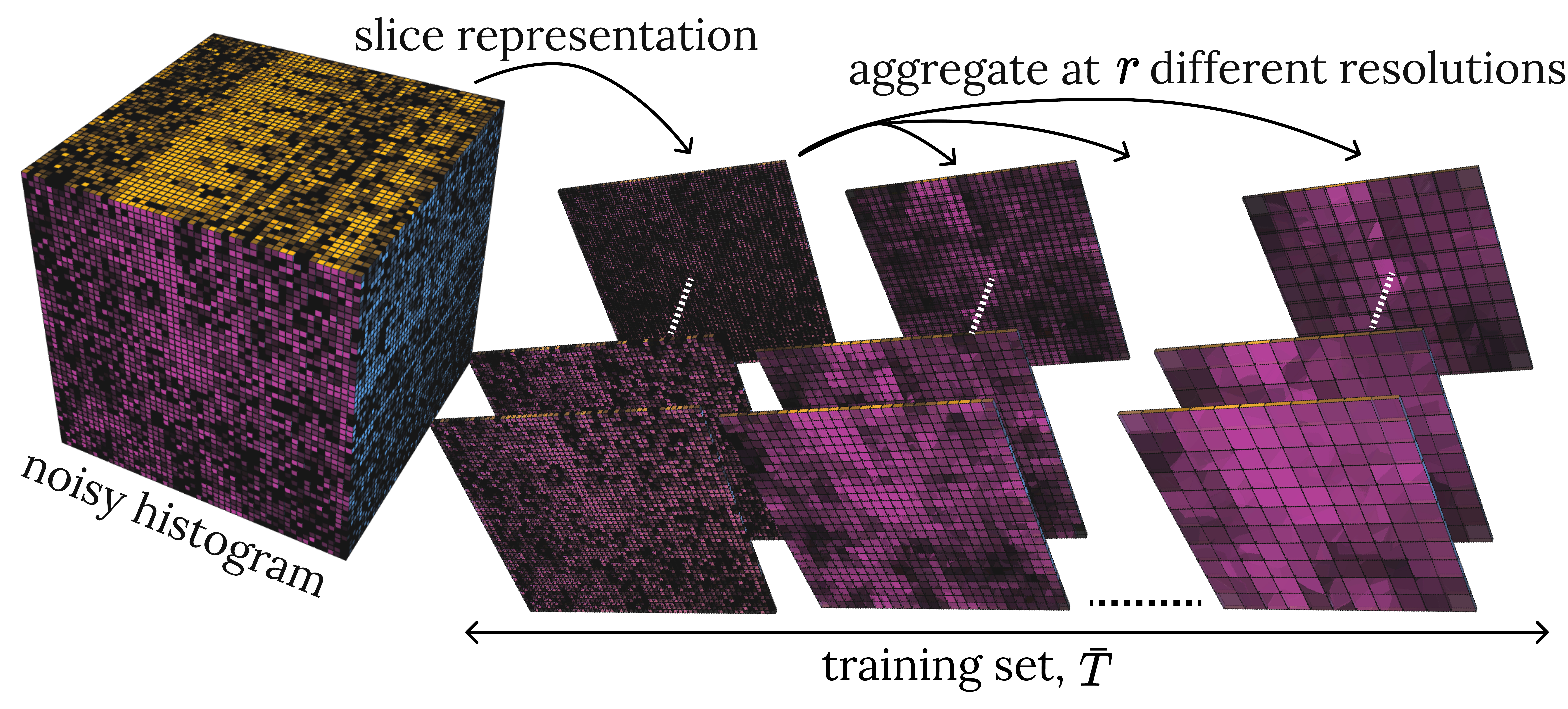}
    \caption{Training set preparation}
    \label{fig:learned_denoising_prep}
\end{figure}
\begin{figure}
    \centering
    \includegraphics[width=\columnwidth]{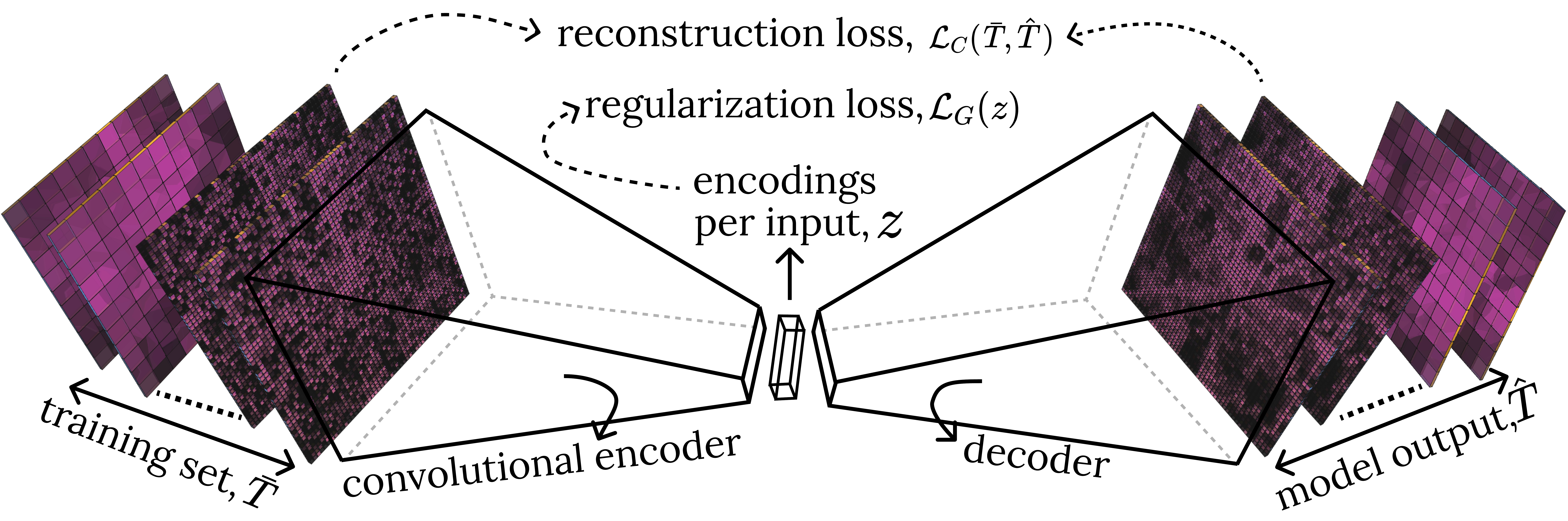}
    \caption{Model Training}
    \label{fig:learned_denoising}
\end{figure}

\subsubsection{Denoising with Convolutional VAEs}\label{sec:decnoising_detail}
Based on the above principles, we utilize convolutional VAEs to denoise the histogram $\bar{H}$. We first provide an overview of our algorithm, then provide more details on the role of regularization in our methodology, and finally present the algorithm pseudocode.

Our method consists of three stages: (1) training data preparation, (2) model training and (3) model inference. We discuss each below. 

\textit{Training data preparation}. Recall that we are given a noisy 3-d histogram, $\bar{H}$, where the 2-d histograms resulting from each slice contain density information for different locations. Thus, we view the 3-dimensional histogram $\bar{H}$ as a set of two dimensional histograms, where the $i$-th element in this set, $\bar{H}_i$, is a 2-dimensional slice corresponding to the $i$-th timestamp. This is shown on the left side of Fig.~\ref{fig:learned_denoising_prep}. Then, as shown on the right side of Fig.~\ref{fig:learned_denoising_prep}, to utilize multi-resolution learning, we aggregate each of the slices at various resolutions. For instance, every block of $2 \times 2$ cells in $\bar{H}_i$ are aggregated to obtain a new 2-d histogram which has a coarser granularity. This aggregation is done at $r$ different resolutions, and all the aggregated histograms are put together in a training set $T$. \revision{We remark that although CNN layers can extract information at multiple resolutions from an input image, multi resolution learning cannot be achieved by adding a CNN layer to the architecture, since doing so would also change the loss function, in order to enforce the information to be captured at multiple resolutions (loss is over all training data i.e., at different resolutions) .}

\textit{Model Training}. We use a convolutional VAE to perform regularized representation learning. An encoder, which is a CNN, takes as input each 2-d slice and outputs a representation for it. Denote by {\tt encoder(.;$\theta_e$)} the network whose parameters are $\theta_e$, and let $\ell$ be the dimensionality of the representation output. The representation is then fed to a decoder, which is another neural network, denoted by {\tt decoder(.;$\theta_d$)}, where $\theta_d$ are the parameters of the decoder. The output of the decoder is a 2-d histogram, as shown in Fig.~\ref{fig:learned_denoising}. To simplify notation, we directly input a set of 2-d histograms to the encoder, in which case the output is also a set of representations (similarly for decoder). The model is trained to minimize a reconstruction loss, which is the difference between input slice and the output slice, and a regularization loss, which ensures that the learned representation follows some regularization constraints. 

\textit{Model Inference}. Each slice, $\bar{H}_i$, is passed through the convolutional VAE, first encoded and then decoded, to obtain the denoised representation for $\bar{H}_i$. This is done for all timestamps, $i$, which allows us to obtain a denoised 3-d histogram, $\hat{{H}}$. Note that, inference is not performed on any aggregated histogram (via multi-resolution), but only on the original noisy histogram $\bar{H}$. In other words, the output of learned denoising is a single 3-d histogram $\hat{{H}}$, which is at the same resolution as the noisy input histogram $\bar{H}$.

\textbf{VAE and Regularization Details}. We discuss parts of the VAE design relevant to the problem of denoising. Further details of our model can be found in Sec.~\ref{sec:exp}.  We utilize the Vector Quantized variant of VAE (VQ-VAE), where the encoding is forced to follow a certain discrete structure \revision{(Sec.~\ref{sec:dicsussion:hyperparams} discusses other possible variants and modelling choices)}. A discrete set, $\Upsilon$, called a \textit{codebook}, of $\mathcal{B}$ different encodings, $\Upsilon=\{e_1, ..., e_{\mathcal{B}}\}$, where each $e_i$ is $\ell$-dimensional, is learned, and VAE training process forces the encoder to output an encoding that is similar to an element in the codebook. Recall that {\tt encoder(.;$\theta_e$)} is a convolutional neural network that takes as input a 2-d histogram. For an input 2-d histogram in our training set, $\bar{T}$, {\tt encoder($\bar{T};\theta_e$)} provides a set of representations $z$. These representations are then input to the decoder to obtain reconstructions $\hat{T}=$ {\tt decoder($z;\theta_d$)}. VQ-VAE defines a distance function between $z$ and $\Upsilon$, $d(z, \Upsilon)$, that measures how similar the encodings are to the codebook. $d(z, \Upsilon)$ is then minimized in the training process to ensure the encoder learns representations that are similar to the codebook. We call $\mathcal{L}_G(z)=d(z, \Upsilon)$ the regularization loss and define $\mathcal{L}_C(\hat{T})=\sum_{i=1}^{|\bar{T}|}\lVert \bar{T}_i-\hat{T}_i\rVert^2$ as the reconstruction loss, where $T_i$ is the $i^{th}$ slice in the training set and $\hat{T}_i$ is the output of the VQ-VAE on the $i^{th}$ training slice. We then train VQ-VAE to minimize $\alpha\times\mathcal{L}_G(z)+\mathcal{L}_C(\hat{T})$, where parameter $\alpha$ is introduced to control the amount of emphasis on the regularization. \revision{We discuss setting the hyperparameters in Sec.~\ref{sec:dicsussion:hyperparams}}.

\textbf{Complete Denoising Algorithm}. The complete denoising process is shown in Alg.~\ref{alg:learned_denoising}. 
Lines~\ref{alg:denoise:create_train}-\ref{alg:denoise:augment} show how the training set is augmented with histograms at varying granularities.  Lines~\ref{alg:denoise:enc}-\ref{alg:denoise:dec} create a CNN as an encoder and a Transposed CNN as the decoder. The model is trained in Lines~\ref{alg:denoise:train_begin}-\ref{alg:denoise:train_end}, where at Line~\ref{alg:denoise:encode} the encoder outputs encodings of the histograms in the training set and the encodings are then decoded by the decoder to reconstruct the histograms in Line~\ref{alg:denoise:decode}. The model is then optimized with stochastic gradient descent to minimize the reconstruction loss and the regularization loss. 
Finally, after convergence, a forward pass of the model yields the denoised histogram. 

We have kept the discussion of convolutional VAEs at a high level and only provided details for ideas that pertain to the problem of denoising, without discussing in-depth the technical details in VAE and VQ-VAE design such as commitment and alignment loss \cite{wu2020vector}, the reparametrization trick \cite{xu2019variance} and their relation to regularization. We provide implementation specific details of VQ-VAE in Sec.~\ref{sec:exp}.

\revision{\subsubsection{\revision{Modelling and Hyperparameter choice}\nopunct}\label{sec:dicsussion:hyperparams}\hfill\\
\revision{\textbf{VAE Type and Architecture}. VAE architectures model data distribution over continuous (e.g., in Gaussian-Process VAE) or discrete (e.g., Vector-Quantized VAE) latent space variables. We limit our discussions to the discrete variant, even though either of these architectures are effective for use in VDR. The regularized representation learning ability of VAEs is essential to VDR, but we do not expect the specific distribution enforced on the latent variables to significantly impact accuracy. Our choice of VQ-VAE is due to its faster training time and convergence rate compared with other variants (see Fig. \ref{fig:gpvae_vs_vqvae} in Sec. \ref{subsec:settings}). Lastly, since we focus on location data that, as discussed in Sec. \ref{sec:data_model}, show homogeneous characteristics across datasets, our choice of VAE architecture is not sensitive to the specific data subset. Thus, we utilize VQ-VAE exactly as in \cite{razavi2019generating} without any modification and across all datasets.} } 

\revision{\textbf{Regularization Parameters}.}
Parameters $\mathcal{B}$ and $\alpha$ control how regularization benefits denoising. (1) $\mathcal{B}$ controls the representation power of the encoding space: the smaller $\mathcal{B}$ is, the less information can be captured by different encodings, as the encodings for different slices are forced to be similar. On the other hand, when $\mathcal{B}$ is large, different slices are allowed to take different representations, as the codebook allows for more variability.  (2) $\alpha$ controls how much the encoder is forced to adhere to the codebook. When $\alpha$ is small, the encoder can learn representations that do not follow the discretized structure. It allows learning different encodings for different slices, thus memorizing the information within slices instead of learning patterns that generalize. 

We find significant benefit in invoking both of the above regularization aspects of VQ-VAEs. Specifically, we saw worse performance when setting $\alpha$ to a small value, confirming our regularized learning design principle and emphasizing the need for regularizing the encoding space. Denoising ability also suffers when $\mathcal{B}$ is too small or too large; the former because not enough information can be stored in the learned encoding, and the latter because the encoding can become noisy (due to insufficient regularization). \revision{Since we focus on location data that, as discussed in Sec. \ref{sec:data_model}, show homogeneous characteristics across datasets, our choice of regularization parameters is not sensitive to the specific data subset. As such, we fixed $\alpha=1$ and $\beta=128$ across all datasets in our experiments.}   

\subsection{Statistical Refinement}\label{sub:refinement}
Given that the values in the denoised histogram are based on the sampled dataset, they will be an underestimation of the true counts. In this section, we study how the values can be scaled to accurately represent the true counts. We first discuss how differential privacy complicates this process of statistical refinement, then present notations and assumptions in our method and finally discuss the statistical refinement step.

\subsubsection{Estimation with Differential Privacy}\label{sec:refine:challanges} Recall that we sampled a dataset $D_s$ from the true dataset $D$, and created a noisy histogram $\bar{H}$ from the sampled set. We retained up to $k$ points per user, hence the size of $D_s$ is smaller than $D$. Thus, the number of data points that fall inside the histogram created based on $D_s$ will be an underestimation of the true number of data points. To adjust the observed answers based on sampled data points we need to scale them, so that they accurately represent the true numbers. However, DP affects how this scaling can be done.

\textbf{Noisy Observations}. 
Scaling the values in $\bar{H}$ scales both the added noise and the observed values, thus amplifying the random noise. In other words, by scaling the observed values, we reduce the bias in our estimation (i.e., account for underestimation), but this scaling increases the variance in our estimation because the random noise gets amplified. Thus, in the case of sampling with differential privacy, it is important for our method to account for both bias and variance in the estimation.

\textbf{Private Sampling Procedure}. The sampling procedure is data dependent, and its specific details may be unknown, due to privacy requirements. Therefore, we aim to derive a refinement approach that is agnostic to the sampling performed during data collection. For instance, the probability of sampling a point in a particular cell does not only depend on the total number of points in that cell, but also on which user they belong to. If all users in a cell have exactly one point in $D$, then all the points in that cell will be preserved and thus the number of points in that cell in $D_s$ will be the same as number of points in the corresponding cell in $D$. However, if users in a cell have more than ${k}$ points, then the number of points in the cell in $D_s$ is less than $D$. Due to differential privacy, information about the number of points per user in a cell can only be known by spending privacy budget, which is undesirable. 

\begin{algorithm}[t]
\begin{algorithmic}[1]
\Require A set of noisy 2-dimensional histograms, $\bar{H}$
\Ensure A set of denoised 2-dimensional histogram, $\hat{H}$
\State $\bar{T}\leftarrow \bar{H}$\label{alg:denoise:create_train}
\For{$j\leftarrow2$ to $r$}
    \For{$\bar{H}_i$ in $\bar{H}$}
        \State $\bar{H}_i^j\leftarrow$ Histogram from aggregating $j\times j$ blocks of $\bar{H}_i$ 
        \State $\bar{T}\leftarrow \bar{T}\cup\bar{H}_i^j$\label{alg:denoise:augment}
    \EndFor
\EndFor
\State {\tt encoder($.;\theta_e$)$\leftarrow$} CNN encoder with params. $\theta_e$\label{alg:denoise:enc}
\State {\tt decoder($.;\theta_d$)$\leftarrow$} TransposedCNN decoder with params. $\theta_d$ \label{alg:denoise:dec}
\Repeat\label{alg:denoise:train_begin}
    \State $z\leftarrow$ {\tt encoder($\bar{T};\theta_e$)}\label{alg:denoise:encode}
    \State $\hat{T}\leftarrow$ {\tt decoder($z;\theta_d$)}\label{alg:denoise:decode}
    \State $\mathcal{L}_C(\hat{T})\leftarrow \sum_i\lVert \bar{T}_i-\hat{T}_i\rVert^2$ \label{alg:denoise:reconstruction_loss}
    \State $\mathcal{L}_G(z)\leftarrow d(z, \Upsilon)$\label{alg:denoise:regularization_loss}
    \State $\mathcal{L}\leftarrow \alpha\times\mathcal{L}_G(z)+\mathcal{L}_C(\hat{T})$ \label{alg:denoise:loss}
    \State $\theta\leftarrow\theta_e\cup\theta_d$
    \State Update $\theta$ in direction $-\nabla_\theta\mathcal{L}$\label{alg:denoise:gd}
\Until{convergence}\label{alg:denoise:train_end}
\State \Return {\tt decoder(encoder($\bar{H};\theta_e)\theta_d$)}\label{alg:denoise:output}
\end{algorithmic}
\caption{Learned Denoising}\label{alg:learned_denoising}
\end{algorithm}

\subsubsection{Estimation algorithm}\label{sec:refine:alg}
Taking into account the above observations, we use mean square error minimization to decide how the answer should be scaled, which accounts for both bias and the variance, and thus ensures that if the noise is too severe it is not amplified. Moreover, rather than spending privacy budget to estimate the sampling procedure, we make simplifying assumptions to create a tractable sampling model that can be mathematically analyzed. In the remainder of this section, we first describe our sampling model and then show how mean square error minimization can be used to decide how the observed noisy answers should be scaled to accurately represent the true data.

\textbf{Notation and Modeling Assumptions}. Let $N=|D|$ be the total number of data points and $n=|D_s|$ be the observed number of data points after sampling. We make simplifying assumptions about the sampling procedure for the purpose of our analysis. Specifically, we consider the case when the $n$ points are sampled independently and uniformly at random. Let $X_i^c$ be the indicator random variable equal to 1 if the $i$-th point falls in a cell $c$. Furthermore, let $\mu_c$ be the proportion of data points in the complete dataset that are in the cell $c$, so that $N\times\mu_c$ is the total number of data points in cell $c$. We assume that the $i$-th point is sampled uniformly at random across all data points, so that $\mathds{P}(X_i^c=1)=\mu_c$. 

\textbf{Algorithm}. Our goal is to design an estimator to estimate $N\times \mu_c$ for all cells, $c$, in the histogram. Our estimator needs to be accurate, but at the same time has to preserve differential privacy. We consider the estimator $\theta_c=\gamma(\sum_iX_i^c+Lap(\frac{k}{\epsilon}))$. $\theta_c$ obtains a differentially private estimate of the observed number of points in the cell $c$ and scales it by a parameter $\gamma$. We find the parameter $\gamma$ by minimizing the mean squared error of our estimator $\theta_c$. To do so we first calculate bias and variance of our estimator.
\begin{align*}
\text{Bias}(\theta_c) &= \mathds{E}[\theta_c-N{\mu}_c]=\mu_c(\gamma n -N)\\
\text{Var}(\theta_c) &= \gamma^2(n\mu_c(1-\mu_c)+2k^2\epsilon^{-2})
\end{align*}
Thus given the mean squared error of an estimator, $\text{MSE}(\theta_c)=\text{Bias}(\theta_c)^2+\text{Var}(\theta_c)$, we obtain 
$$
\text{MSE}(\theta_c)=\gamma^2(n\mu_c(1-\mu_c)+2k^2\epsilon^{-2})+\mu_c^2(\gamma n -N)^2.
$$

Next, we find the $\gamma$ value that minimizes error across all cells. Let $m=$$\mathbb{M}$$\times$$\mathbb{M}$$\times$$\mathbb{T}$ be the number of cells in the histogram. We minimize $\sum_{c=1}^m\text{MSE}(\theta_c)$ by taking the derivative of $\sum_{c=1}^m\text{MSE}(\theta_c)$ with respect to $\gamma$ and setting it to zero. We obtain that 
\begin{align}\label{eq:gamma}
    \gamma =\frac{nNC}{2mk^2\epsilon^{-2}+(1-C)n +Cn^2}
\end{align}
minimizes $\sum_{c=1}^m\text{MSE}(\theta_c)$, where $C=\sum_{c=1}^m\mu_c^2$ is a data-dependent constant. 
It is left to determine the value of $C$, but doing so on the private data itself may require spending privacy budget. However, due to inherent properties of location datasets, in practice, $C$ can be treated as a system parameter and set---in a data-independent manner thus not consuming valuable privacy budget---to a fixed value that works. We further discuss how $C$ can be set in Sec.~\ref{sec:system:trade-offs}.

\section{System Design and Analysis}\label{subsec:system}
\begin{algorithm}[t]
\begin{algorithmic}[1]
\Require A dataset $D$, privacy budget $\epsilon$, spatial, $\mathbb{M}$, and temporal, $\mathbb{T}$, discretization granularity, sampling parameter $k$, and refinement factor $C$. 
\Ensure Differentially private 3d-histogram $\hat{H}$ of $D$
\State $D_s\leftarrow$ sample $k$ points per user in $D$\label{alg:final:sample}
\State $H_s\leftarrow$ $\mathbb{M}$$\times$$\mathbb{M}$$\times$$\mathbb{T}$ histogram of $D_s$\label{alg:final:discritize}
\State $\bar{H}\leftarrow H_s+Lap(\frac{k}{\epsilon})$\label{alg:final:noise}
\State $\hat{H}\leftarrow \text{Denoise}(\bar{H})$\label{alg:final:denoise}
\State \Return $\gamma_C\times \hat{H}$\label{alg:final:upsample}
\end{algorithmic}
\caption{\name{} algorithm}\label{alg:final_alg}
\end{algorithm}

\subsection{Privacy Analysis}\label{subsec:priv_anal}
Alg.~\ref{alg:final_alg} shows our proposed end-to-end algorithm. Lines~\ref{alg:final:sample}-\ref{alg:final:noise} correspond to the data collection step, line~\ref{alg:final:denoise} calls Alg.~\ref{alg:learned_denoising} to perform learned denoising and line~\ref{alg:final:upsample} uses the value of $\gamma_C$ calculated in Eq.~\ref{eq:gamma} to scale the results (we write it as $\gamma_C$ to make explicit the dependence on the factor $C$). Alg.~\ref{alg:final_alg} only accesses the data in the data collection step. Thus, lines~\ref{alg:final:denoise}-\ref{alg:final:upsample} can be seen as a post processing step and do not consume any privacy budget. Data collection is $\epsilon$-DP, and thus the entire \name{} provides $\varepsilon$-DP, as proved by the following:

\begin{theorem}
Algorithm~\ref{alg:final_alg} is $\epsilon$-DP.
\end{theorem}
\revision{The proof builds upon \cite{dwork2014algorithmic,kellaris2013practical} (privacy guarantee of our sampling approach follows Theorem 1 of \cite{kellaris2013practical})} and is available in our extended technical report \cite{vdr_technical}.


\subsection{\name{} Design Choices}\label{sec:query_discussions}
Real world use-case of spatio-temporal data extends beyond simple range count queries that are commonly studied and optimized-for in common approaches to location data release, which are typically partitioning-based \cite{qardaji2013differentially, qardaji2013differentially, zhang2018ektelo,qardaji2013understanding, zhang2017privbayes, zhang2014towards, hardt2012simple}. Other common query types such as forecasting POI visits, or finding hotspots, are sensitive to biases that such approaches introduce, causing them to perform poorly. VDR's approach of denoising a histogram created by Laplace mechanism (LPM) offers significant benefits across different spatio-temporal queries by avoiding implicit biases. 

\textbf{Forecasting Queries}. Forecasting methods are often robust to random noise present in real data, some even explicitly incorporating its effects in their models. Thus, a DP mechanism that only introduces random noise, such as LPM, can perform well, whereas those that approximate the density of regions by cleverly grouping and partitioning the domain introduce additional bias and obliterate trends and seasonal effects present in the timeseries.

\textbf{Hotspot Queries}. If a DP mechanism underestimates counts in the region of a hotspot, it will receive a distance penalty due to not having found the correct spot, and may incur a large {\em regret}, up to the maximum density threshold. This happens if an approach creates coarse partitions of the data, thus underestimating the density for `hot' peaks. Selectively creating finer partitions can improve the result, since some `hot' peaks may be preserved. Nonetheless, modelling errors in deciding where to create fine partitions can cause underestimation in some regions, resulting in large distance penalty. On the other hand, a bias-free approach such as LPM performs better since it is not affected by a systematic reduction in data utility that partitioning approaches incur. 

\textbf{Range Count Queries}. Answering larger query ranges over LPM requires aggregating more histogram cells, each contributing additional error to the answer. Therefore, VDR is specifically designed to denoise (reduce variance) of a bias-free mechanism, smoothing out the noise by exploiting the inductive bias that spatial patterns exist in location datasets. In this way, it can improve forecasts significantly by preserving timeseries specific factors and discovering hotspots that likely meet the threshold, while not sacrificing the quality of results for range count queries.

\revision{\textbf{Applicability to Other Queries}. We focus on typical spatiotemporal queries that avoid release of user-level data in order to be consistent with differential privacy objectives. Thus, our approach is specifically useful for queries that require and/or ask for patterns within the data, as regularized learning is designed to reduce noise by learning spatial patterns in the data. However, such an approach is not suitable if a query asks for data nuances as they may not be captured along with the learned patterns, or if the query asks for user specific information (e.g., distribution of locations per user), since such leakage is in conflict with differential privacy objectives.}

\subsection{System Parameter Selection}\label{sec:system:trade-offs}
We discuss the impact of system parameters $k$, $C$ and data collection granularity on the performance of the system, and provide guidelines on how they should be set in practice.

\subsubsection{Refinement Factor and Sampling Parameter}\label{sec:sampling_parameters}
Recall that in our data collection step (Alg.~\ref{alg:final_alg} line~\ref{alg:final:sample}), we sample up to $k$ points per user and in the statistical refinement step we scale our result by a factor $\gamma_C$ (Alg.~\ref{alg:final_alg} line~\ref{alg:final:upsample}), which depends on the refinement factor $C$. Both parameters, as discussed below, depend on data skewness as well as distribution of user contributions. However, due to DP, measuring data dependent properties requires spending privacy budget, which is scarce. 
Next, we discuss potential trade-offs in the values of these system parameters, and heuristics to set each.

\textbf{Sampling parameter, $k$.} For accuracy, the sampled dataset should retain density characteristics of the original dataset. After scaling with our statistical refinement step, the obtained query answers should be close to original counts. In our real-world datasets, the true data size, $N$, plays an important role in the interplay between true data characteristics and sampled ones. Specifically, when $N$ increases, most cells in the true 3-d histogram, $H$, remain empty or retain small values, due to data sparsity, while the number of reported locations in dense cells increases. This results in a more skewed true dataset. Thus, for the sampled dataset $D_s$ to capture this skewness, we need a larger number of samples, or otherwise our estimation will have a very large variance. We conclude that the value of $k$ should grow with data size. Our experiments in Section~\ref{sec:exp:systemanalysis} corroborate this heuristic, showing that the growth ratio $\lambda$, defined as $\frac{k^*}{N}$,
where $k^*$ is the best possible sampling rate, stays almost constant across datasets of various sizes \revision{(see Fig.~\ref{fig:ulvl_growthratio})}. In fact, we observe that this value remains constant across different cities, suggesting that due to similarity in density patterns   inherent to location datasets \revision{(as discussed in Sec.~\ref{sec:data_model})}, we can set the value of $k$ to be a constant fraction of $N$. Details of these observations are presented in Sec.~\ref{sec:exp:ulvl}.

\textbf{Refinement Factor, $C$.} Recall that $C$ determines how query answers are scaled to obtain the final histogram. Our theoretical model suggests that $C=\sum_c \mu_c^2$, but we do not have access to $\mu_c$ due to differential privacy, thus we treat $C$ as a system parameter. Note that, $C$ depends on data distribution. For instance, if the data are uniformly distributed, $\mu_c$ will be equal to $\frac{N/m}{N}=\frac{1}{m}$, so that $\sum_c \mu_c^2=m\times\frac{1}{m^2}=\frac{1}{m}$. On the other hand, if all the $N$ points are in a single cell, then $\sum_c \mu_c^2=1$. For location datasets, we expect similar density patterns across the space \revision{(as discussed in Sec.~\ref{sec:data_model})}, and as a result, similar values of $C$ should perform well across datasets. Our results in Sec.~\ref{sec:exp:ulvl} confirm that the same value of $C$ can be used with distinct datasets and sampling rates \revision{(see Fig.~\ref{fig:vdr_vs_debiasingC})}. We suggest setting the value of $C$ to one that performs well on a public dataset. Having $C$ as a system parameter is advantageous because it allows for correction of errors that have been introduced due to our theoretical modeling. For instance, our analysis in Sec.~\ref{sec:refine:alg} does not take into account the impact of denoising, which we expect to be consistent across datasets. By setting $C$ as a system parameter, we can avoid any adverse impacts of modeling errors in practice. Moreover, since the modelling error is consistent across datasets, the same value of $C$ can be set for all datasets.

\subsubsection{\revision{Spatio-temporal Resolution}}\label{sec:spatiotemproal_resolution}
\revision{We assume throughout our discussion that a 3-d histogram of a predefined resolution is required, which is often the case since the choice of resolution is domain specific. High-resolution density maps are preferred for location datasets in industrial data release projects~\cite{dataforgood, aktay2020google}. VDR uses a 30m$\times$30m grid and 3h temporal resolution. A coarser discretization induces partitioning biases in query answers: the answer to finer queries are estimated from the answer to their enclosing coarse cell assuming uniformly distributed points within the spatial extent (Sec.~\ref{sec:query_discussions} discusses the adverse effects of such a bias). 
Nonetheless, there are limits to how high the resolution of the data release can be for two reasons: (1) A fine-grained histogram will have small true count values per cell, and since scale of DP-added noise is proportional to sensitivity and not the counts, the resulting signal-to-noise ratio will be low, compromising accuracy. 
(2) The spatio-temporal resolution needs to pick up the existence of consistent spatial patterns. VDR's accuracy may suffer if too fine of a granularity is chosen, as no consistent spatial patterns may exist for the CNN layers to learn. 
We experimentally support these conjectures in Fig. \ref{fig:temporal_resolution} in Sec.~\ref{subsec:system_modelling}.}

\subsection{Impact of Assumptions}\label{sec:assumptions_impact}


\revision{Specific assumptions on spatial-data distribution (as discussed in Sec~\ref{sec:data_model}) have moulded VDR design (as discussed in \ref{sec:query_discussions}). Consequently, a discussion is warranted on the effect on VDR when these assumptions do not hold:
(1) The use of {\em CNNs} is predicated on the existence of repeating {\em spatial} patterns within a city. If such an assumption does not hold, model accuracy suffers. For instance, as data become more uniformly distributed, the accuracy gain of VDR compared to benchmarks is lost. However, in practice, location data are often skewed, and the release of uniform-like data is a far less challenging/interesting problem for which simpler solutions may suffice.
(2) To avoid data-dependent parameter tuning, VDR relies on the existence of similar density patterns {\em across} cities. If this assumption fails (e.g., when considering the release of generic 3-d dataset) one may need to spend additional privacy budget on system and hyper-parameter tuning, hence overall accuracy will decrease. (3) Due to the power law distribution discussed in Sec.~\ref{sec:data_model}, our sampling strategy captures the bulk of density information for small values of $k$, thus lowering query sensitivity and improving the signal-to-noise ratio (also see Sec.~\ref{sub:collection}). However, if user contributions are more uniform (e.g., all users have the same number of contributions), sampling (or the subsequent refinement) may not be effective, and keeping all user data may acheive a better signal-to-noise ratio.}



\vspace{-2mm}
\subsection{Data Release over Time} So far we have considered the release of a static dataset $D$. In practice, spatiotemporal data is released over time, with new data coming in continuously. In such a setting, privacy budget is often allocated per time period, e.g., a budget of $\varepsilon_i$ would be allocated for the $i$-th week ($\varepsilon_i$ is typical set to go to zero so that $\sum_i\varepsilon_i$ is bounded). Thus, the release consists of a sequence of datasets $D_1, D_2, ...$, where each $D_i$ covers a fixed period of time. Let  $\tau$ denote the duration covered by each $D_i$, which we call release duration. To use \name{} in this setting, Alg.~\ref{alg:final_alg}, can be called for every release, where in the $i$-th release, the input dataset is $D_i$ and privacy budget is $\epsilon_i$. However, an important characteristic of \name{} is that the model does not need to be retrained for every data release. That is, rather than retraining the model in the learned denoising step for every release, after the model is trained once, it is still able to denoise the input histograms. We verify this empirically in our experiments. This also shows that our model is learning recurring patterns from data, which generalize well to unseen data points.

\section{Experimental Evaluation}\label{sec:exp}


\subsection{Experimental Settings}
\label{subsec:settings}


\noindent\textbf{Datasets.} 
User check-ins are specified as tuples of: user identifier, latitude and longitude of check-in location, and timestamp. Our primary dataset is the proprietary {\em Veraset}~\cite{verasetref} data {\em(VS)}, a data-as-a-service company that collects anonymized movement data from 10\% of the cellphones in USA \cite{verasetcoverage}. For a single day in Jan 2020, there were 2.4 billion readings from 27.2 million distinct users. 

We also present results on public datasets, containing sporadic check-ins made over a relatively long period of time, as opposed to real longitudinal trajectories of users that the proprietary dataset offers. The {\em Foursquare} dataset {\em (4SQ)}~\cite{yang2019revisiting} is collected during a period of $22$ months and has 22M checkins by 114k users at 3.8 M POIs. The {\em Gowalla (GW)} dataset from the SNAP project~\cite{cho2011friendship} contains 6.4 million records from 196k unique users between February 2009 and October 2010. The {\em San Francisco taxi} dataset {\em (CABS\_SF)}~\cite{piorkowski2009crawdad} is derived from the GPS coordinates of approximately 500 taxis collected over 24 days in May 2008. 

We consider urban areas in the US covering 20km$ \times $20km each. For the Veraset data, we select cities based on their population density~\cite{pop_density}. We selected Salt Lake City, UT (VS\_SL) as a \textit{low density} city (41M points from 600k users), Los Angeles, CA (VS\_LA) as \textit{medium density} city (80M points from 852k users), and Houston, TX (VS\_HT) as high density city (221M points from 1.28M users). For all primary datasets, we discretize the temporal domain to 3 hours, giving a total of $\mathbb{T}=96$ slices for the 12 day period from Jan 7 to 19, 2020. For the secondary datasets, we discretize the temporal dimension such that each slice covers the duration of one month for a total of $\mathbb{T}=24$ slices. From the Foursquare dataset we consider its Tokyo, Japan subset (4SQ\_TKY) with 755k location updates from 8k unique users. From Gowalla we select the San Francisco (GW\_SF) subset with 568k location updates from 14k users and New York (GW\_NY) city with 520k location updates from 16k users. For the CABS dataset, following \cite{hay2016principled,qardaji2013differentially}, we keep only the start and end points of the mobility traces, for a total of 846k records. 

\noindent\revision{\textbf{Dataset Characteristics}. In all our datasets, we observed the power law distribution across users as discussed in Sec.~\ref{sec:data_model}. Summary power law statistics and visualizations are presented in Sec.~\ref{appx:data_charactaristics}. To quantify the existence of temporal patterns, we extracted density timeseries for each of the $\mathbb{M}\times\mathbb{M}$ cells in the histogram 
and evaluated an Autocorrelation Function (ACF) on the  timeseries. We found that only 6.2\% (VS\_HT), 2.4\% (VS\_LA), and 1.5\% (VS\_LA) of such series exhibit non-stationarity, showing that for most cells, we do not observe such temporal patterns. }

\noindent\textbf{Parameter Settings.}
Following \cite{dataforgood,li2014data, zhang2016privtree}, we partition the space into a $576$$\times$$576$ (i.e., $\mathbb{M}=576$) grid to obtain $30m\times30m$ cells. As described above, the temporal granularity is specific to each dataset. The default value of privacy budget $\varepsilon$ is set to $0.2$.

\noindent\textbf{Evaluation Metrics.} For range count queries, we construct query sets of 5,000 RCQs centered at randomly selected data records. Each query has side length that varies uniformly from 30 meters to 120 meters. We set smoothing factor $\psi$ to 0.1\% of the cardinality $n/\mathbb{T}$ of the average time slice of the spatio-temporal dataset~\cite{zhang2016privtree,cormode2012differentially,qardaji2013differentially}. When comparing multiple datasets with each other, the smallest smoothing constant among them is used to remain consistent. 

We evaluate forecasting queries on Veraset subsets only, since other datasets do not contain timeseries of sufficient length. \revision{In order to find forecastable timeseries in the majority of queries, we sampled positions of POIs in the city and extracted timeseries of random lengths that satisfy a ACF Seasonality test~\cite{makridakis2000m3} (90\% confidence) at seasonal period of 8 (meaning a daily seasonality according to 3 hour temporal discretization of the 24 hour period)}.
To make forecasts, we use the winning algorithm of the M3 forecasting competition~\cite{makridakis2000m3}, Theta~\cite{assimakopoulos2000theta}, which is a variant of the Simple Exponential Smoother.
We use the data of all-but-last day to fit the forecaster and evaluate its predictions for last day (i.e., a horizon of $h=8$). We report the sMAPE error (Sec.~\ref{sec:bck}) on 100 forecastable timeseries.

Lastly, the hotspot query is evaluated at a specified threshold $\nu$. Queries originating at random from 1000 randomly selected data points are answered using an expanding search within the 3-d spatial region $SR$ with lengths 5km in spatial extent and no bounds in the time dimension.

\noindent\textbf{Implementation.}  All algorithms were implemented in Python, and executed on Linux machines with Intel i9-9980XE CPU, 128GB RAM and a RTX3090 GPU. Neural networks are implemented in JAX~\cite{jax2018github}. Given this setup, VDR took up to 50 minutes to train for 12 days of the Veraset Houston data. The inference time of VDR is less than 1ms and the model takes 9 MB of space. We publicly release the source code at~\cite{vdr_implementation}.

\begin{figure}
    \begin{minipage}[t]{0.45\linewidth}
        \includegraphics[width=\linewidth]{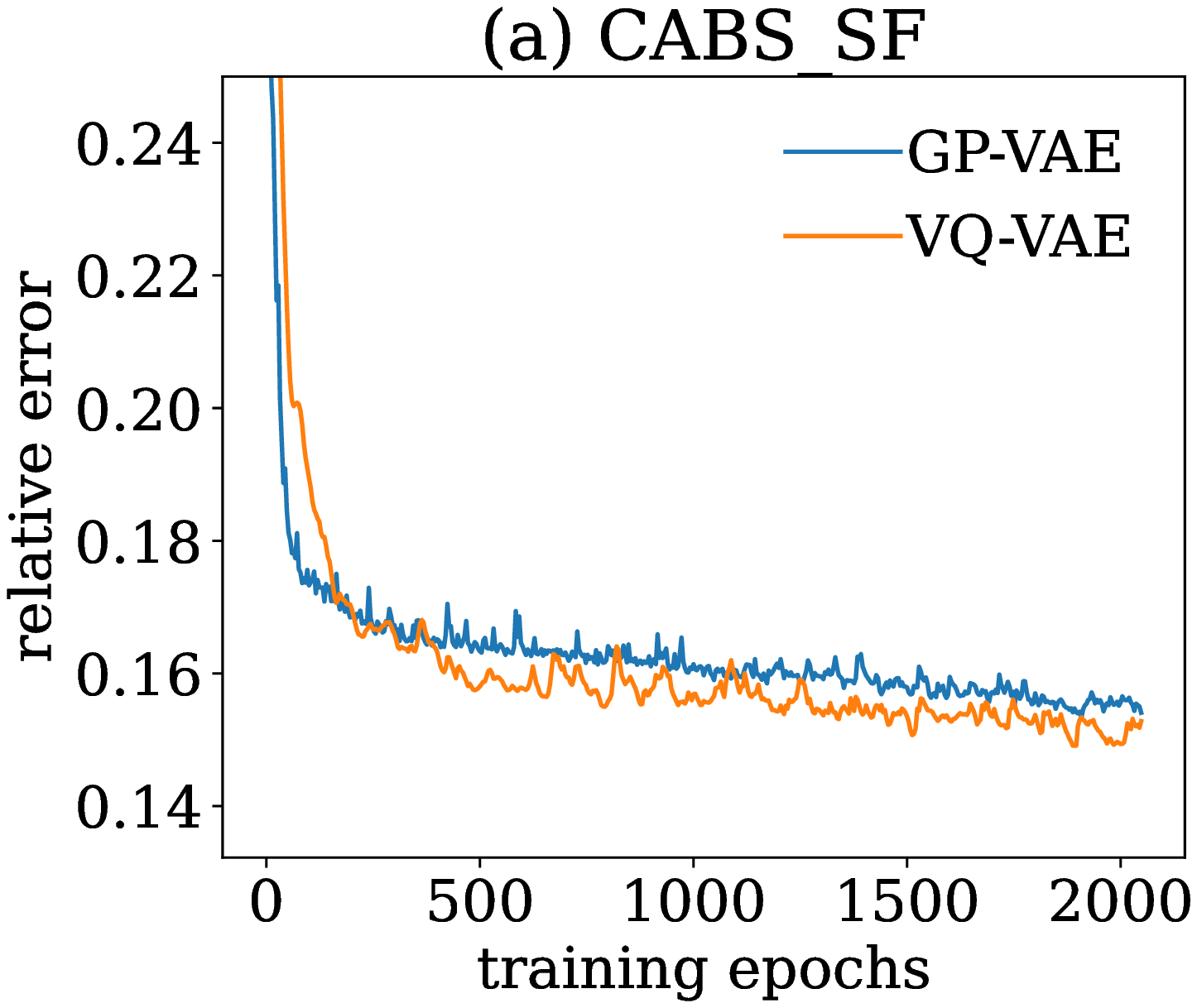}
    \end{minipage}
    \begin{minipage}[t]{0.45\linewidth}
        \includegraphics[width=\linewidth]{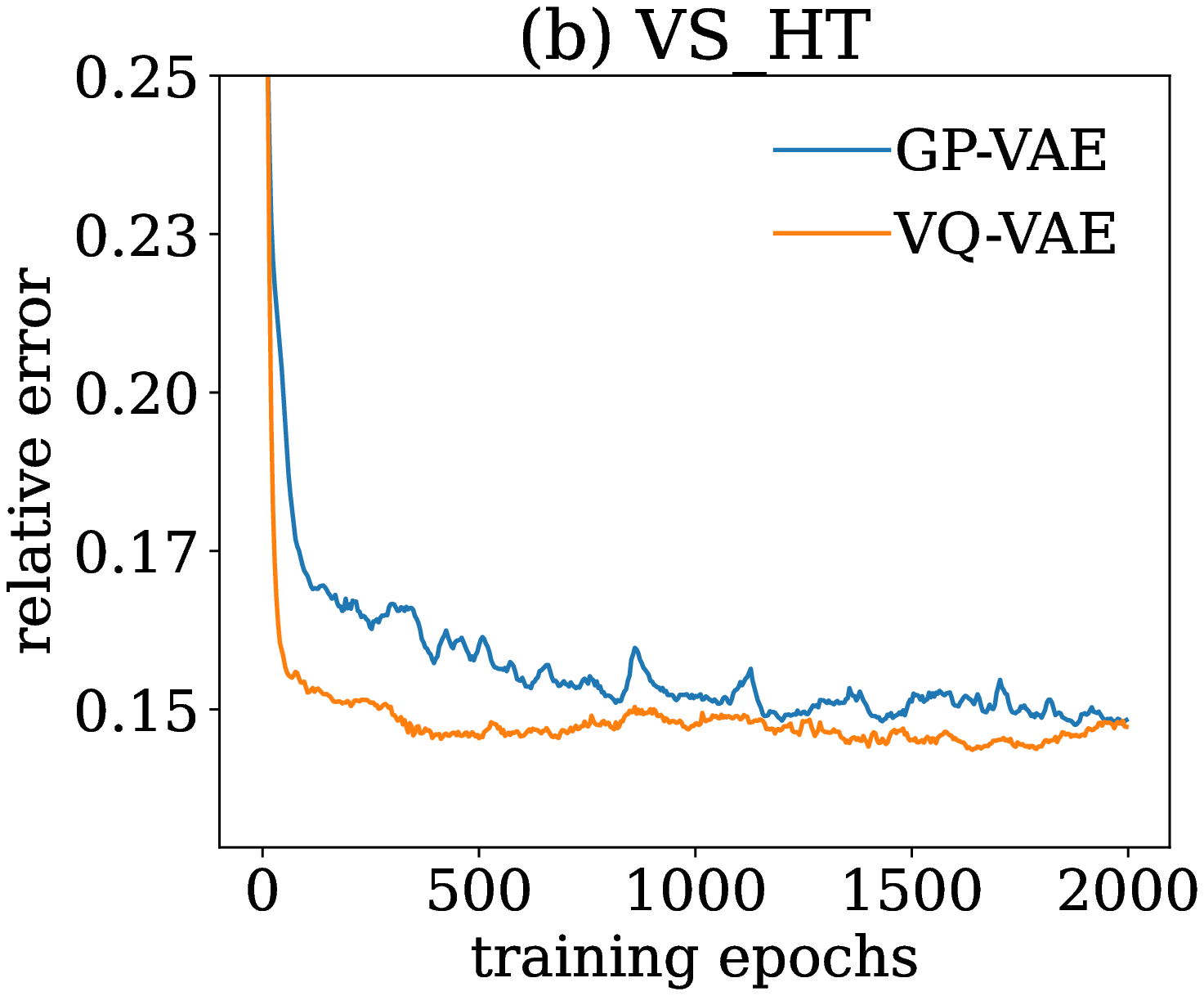}
    \end{minipage}   
    \caption{Choice of Variational AutoEncoder}
    \label{fig:gpvae_vs_vqvae}
    \vspace{-5pt}
\end{figure}

\noindent\textbf{Model Training.} For Multi-Resolution Learning we augment the training set at $r=3$ granularities chosen at equal spacing between the minimum (30m) and maximum (120m) query ranges to be evaluated. The encoder and decoder architecture is based on ResNet\cite{targ2016resnet}. The model takes as input batches of histogram slices and passes them through the ConvNet encoder $\theta_e$ and decoder $\theta_d$. 
\revision{The choice of VAE is primarily motivated by faster training times and often better accuracy for datasets with multiple slices. In Figure~\ref{fig:gpvae_vs_vqvae} we compare a Gaussian-Process VAE (which learns a continuous latent variable) to the Vector-Quantized VAE as used in VDR (which learns a discrete latent variable, i.e., a codebook). For the models to achieve their lowest error, we observe that VQ-VAE requires a lower number of epochs on both public (CABS\_SF) and proprietary (VS\_HT) datasets. The gap is notable in VS datasets due to them having a larger number of slices to train on, which are in-addition multiplied by Multi-Resolution Learning. In such instances, VQ-VAE is able to both train quicker and achieve better accuracy. To train the VQ-VAE, we utilize the Adam~\cite{kingma2014adam} optimizer with Exponential Moving Average updates \cite{razavi2019generating}. The EMA version trained much faster than the non-EMA version, especially when using MRL. Using EMA updates also has the advantage that the embedding updates are independent of the choice of optimizer used for the encoder, decoder and other parts of the architecture.} 
\revision{The loss function used to train the codebook is as in \cite{van2017neural}}.
 In all our experiments, we utilize hyperparameters consistent with those utilized in previous work~\cite{van2017neural,razavi2019generating}; i.e., $\ell=64$, $\mathcal{B}=128$ and a batch size of $b=8$. \revision{ $\alpha$ is used only to study the impact of regularization on model robustness and is set to 1 (equal weighting).}

\vspace{-2pt}

\subsection{Comparison with Baselines} \label{sec:exp:comparison}
\noindent\textbf{Baselines.}
We use as benchmarks  Uniform Grid (UG) \cite{qardaji2013differentially}, Adaptive Grid (AG) \cite{qardaji2013differentially}, HB\_striped \cite{zhang2018ektelo,qardaji2013understanding}, PrivBayes~\cite{zhang2017privbayes}, AHP~\cite{zhang2014towards} and MWEM~\cite{hardt2012simple}. Brief summaries of each method are provided in Sec.~\ref{sec:rel_works}. We utilize Ektelo \cite{zhang2018ektelo,ektelogit}, an operator-based framework for implementing privacy algorithms. To extend approaches designed to originally support range queries in two-dimensional data (spatial-only) to the 3-d case, we partition the temporal domain into non-overlapping slices, so that the measurements are essentially the 2D histograms resulting from each slice. For example, HB\_striped\cite{zhang2018ektelo} performs on each slice the HB algorithm \cite{qardaji2013understanding}, which builds an optimized hierarchical set of queries. We similarly implement Uniform Grid (UG)\cite{qardaji2013differentially} and Adaptive Grid (AG)\cite{qardaji2013differentially}. We use as-is algorithms that are designed to extend to the multi-dimensional setting such as PrivBayes~\cite{zhang2017privbayes}, AHP~\cite{zhang2014towards} and MWEM~\cite{hardt2012simple}. 
We were unable to run the DAWA \cite{li2014data} algorithm directly on such a large domain due to memory and computational constraints. DAWA is designed for 1D-inputs and extended to 2D using a Hilbert fractal. 

\textbf{Privacy Model}. Since none of the baselines consider user-level privacy, to allow for a fair comparison we present experiments with event-level privacy. We disable VDR's sampling and statistical refinement steps, and assume each record belongs to a separate user, hence the privacy protection offered degrades to event-level. We evaluate VDR with user-level privacy in Sec.~\ref{sec:exp:ulvl}.

\vspace{-8pt}
\subsubsection{Range Count Query}

\begin{figure*}
    \begin{minipage}[t]{\textwidth}
        \includegraphics[width=\textwidth]{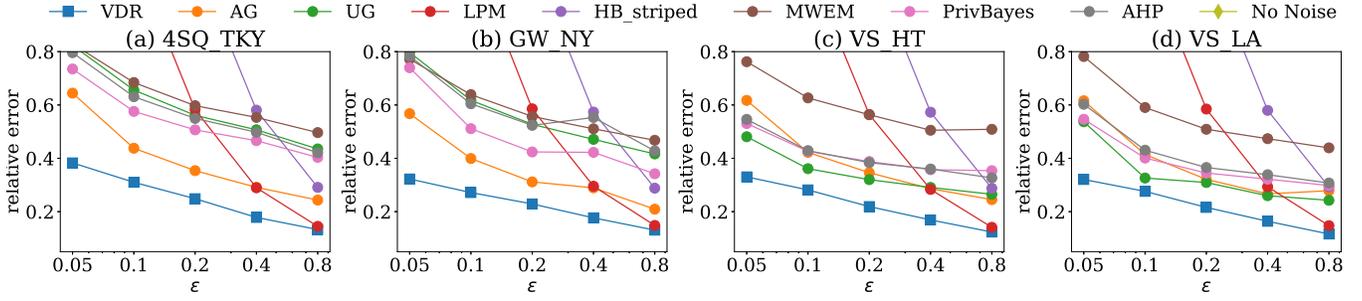}
        \vspace{-5mm}
        \caption{Impact of privacy budget on range count query (RCQ) accuracy.}
        \label{fig:rcq}
    \end{minipage}  
    \vspace{-5mm}
\end{figure*}

\begin{figure*}
    \begin{minipage}[t]{0.375\textwidth}
        \includegraphics[width=\textwidth]{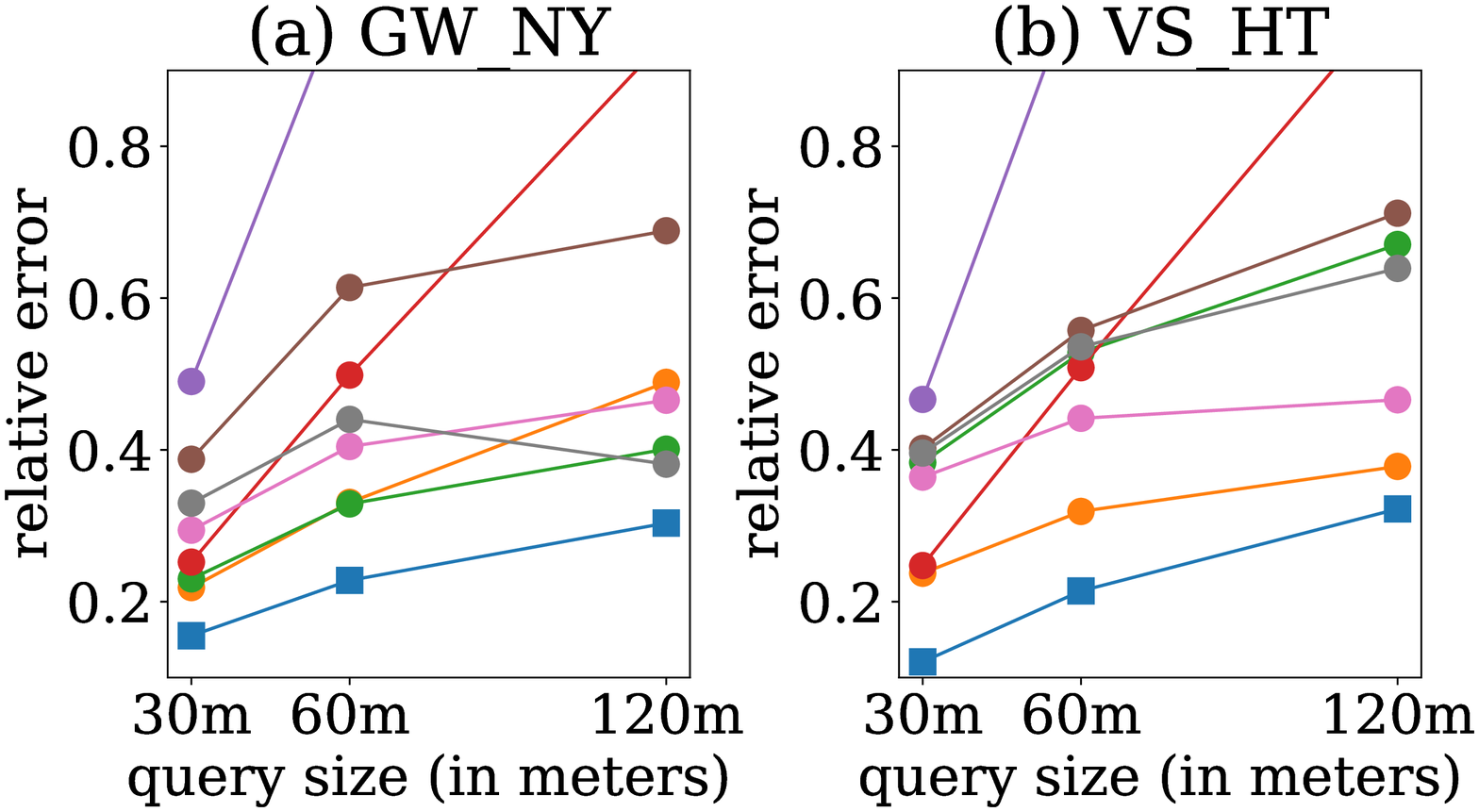}
        \vspace{-3mm}
        \caption{Impact of query size on RCQ.}
        \label{fig:qsize}
    \end{minipage}
    \hfill
    \begin{minipage}[t]{0.615\textwidth}
        \includegraphics[width=\textwidth]{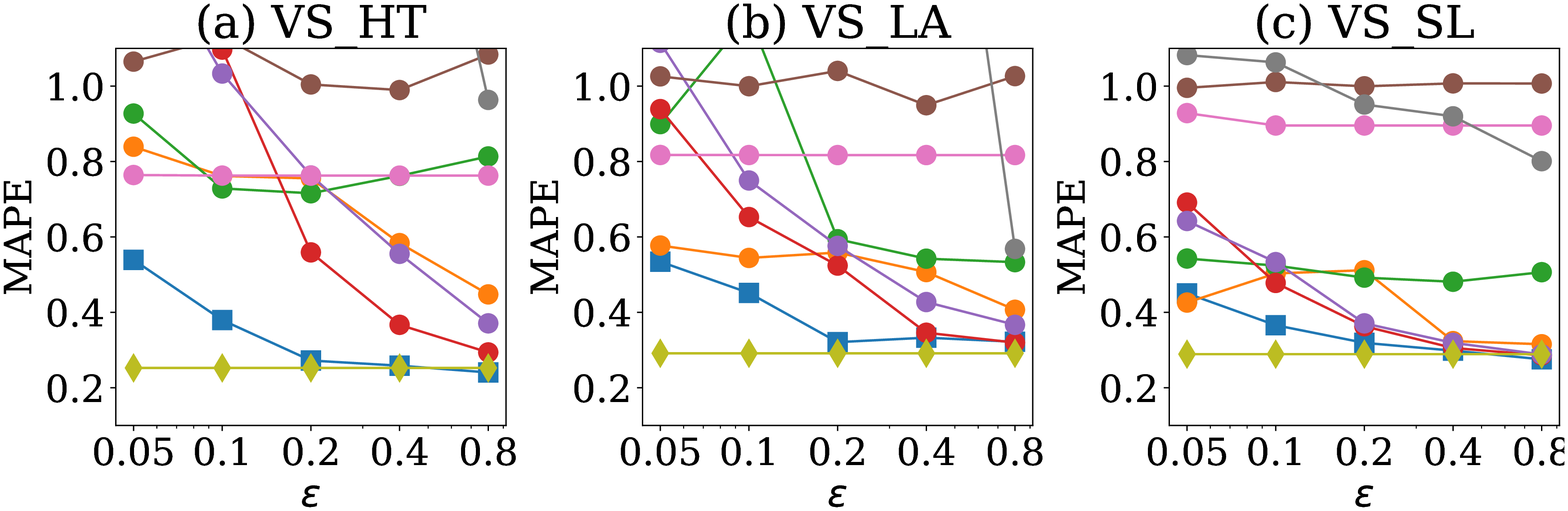}
        \vspace{-3mm}
        \caption{Impact of privacy budget on  forecast query error (sMAPE)}
        \label{fig:fcast}
    \end{minipage}  
    \vspace{-5mm}
\end{figure*}

\begin{figure*}
    \begin{minipage}[t]{\textwidth}
        \includegraphics[width=\textwidth]{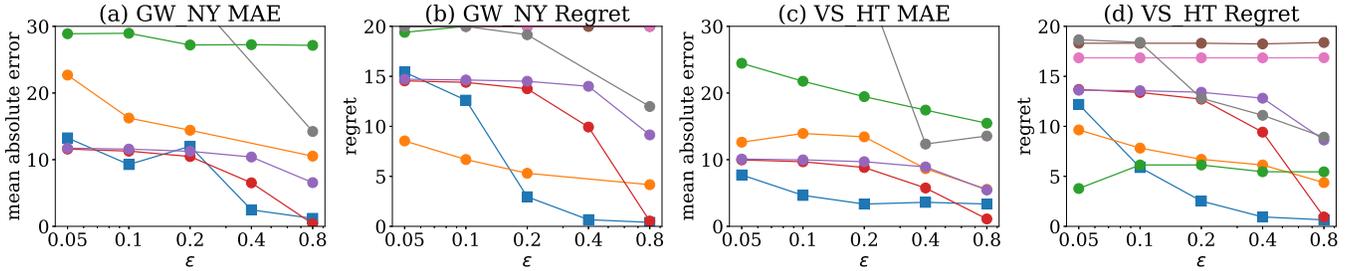}
    \vspace{-5mm}
    \caption{Impact of privacy budget on hotspot query accuracy.}
    \label{fig:hotspot_threshold20}
    \end{minipage}  
    \vspace{-2mm}
\end{figure*}

Figure ~\ref{fig:rcq} presents the error of VDR and the compared approaches when varying $\varepsilon$ (smaller $\varepsilon$ means stronger privacy protection). Unsurprisingly, the error reduces as $\varepsilon$ increases, with VDR consistently outperforming all competitor approaches. VDR is effective in capturing spatial patterns from the data and using them to smooth excessive noise. 

Figure~\ref{fig:qsize} evaluates the impact of query size on accuracy for datasets GW\_NY and VS\_HT (similar trends were observed for other datasets). The error increases when query size grows, due to the fact that computing the result requires aggregating more grid cells, each contributing additional error to the answer. Still, VDR consistently outperforms competitors at all query sizes.

\subsubsection{Forecasting Query}
Figs.~\ref{fig:fcast}(a)-(c) show results on Veraset data subsets VS\_HT, VS\_LA and VS\_SL (corresponding to high, medium and low density). Remarkably, for $\varepsilon=0.2$ and higher, VDR performs as well as the non-private benchmark (`No Noise'). VDR is robust to density changes in the data, as its performance is not significantly different across the three datasets.
VDR significantly outperforms all private benchmarks, with its ability to smooth out the noise. In some instances, e.g., $\varepsilon=0.05$ of Fig~\ref{fig:fcast}(c), UG can perform well, but it does so by making \textit{naive} forecasts that predict the last period's actuals as next period's value, without establishing causal factors.

\vspace{-3pt}

\subsubsection{Hotspot Query}
Fig.\ref{fig:hotspot_threshold20} reports the accuracy for hotspot queries on various Veraset subsets at the fixed density threshold of $\nu=20$. We report the results for varying thresholds in Sec.~\ref{appx:exp} of our technical report~\cite{vdr_technical}.
Mechanisms that model directly the data distribution such as MWEM, AHP and PrivBayes tend to underestimate density globally, and incur a large MAE and regret penalty, up to the maximum of the density threshold. To the same effect, UG, due to its coarse partitioning of the data domain, underestimates the `hot' peaks that the query searches for, also experiencing both a large MAE and regret. AG improves these estimates to some extent by building a finer domain partitioning in the lower level of its hierarchy, and while it may not locate the closest hotspot (high MAE), it still finds one that meets the density threshold (lower regret). LPM fares well for this query as it is not affected by the biases that partitioning approaches bring about. VDR further improves on LPM in both metrics.  In all instances, VDR finds an effective balance between the MAE and regret error metrics.

By starting with an unbiased estimate of the density counts and denoising them, VDR clearly outperforms the existing state-of-the-art in all settings. Next, we no longer consider competitor approaches, and we focus on analyzing the behavior of VDR when varying system parameters.

\subsection{System Analysis}\label{sec:exp:systemanalysis}
\begin{figure*}
    \centering
        \begin{minipage}[t]{0.23\textwidth}
        \includegraphics[width=\textwidth]{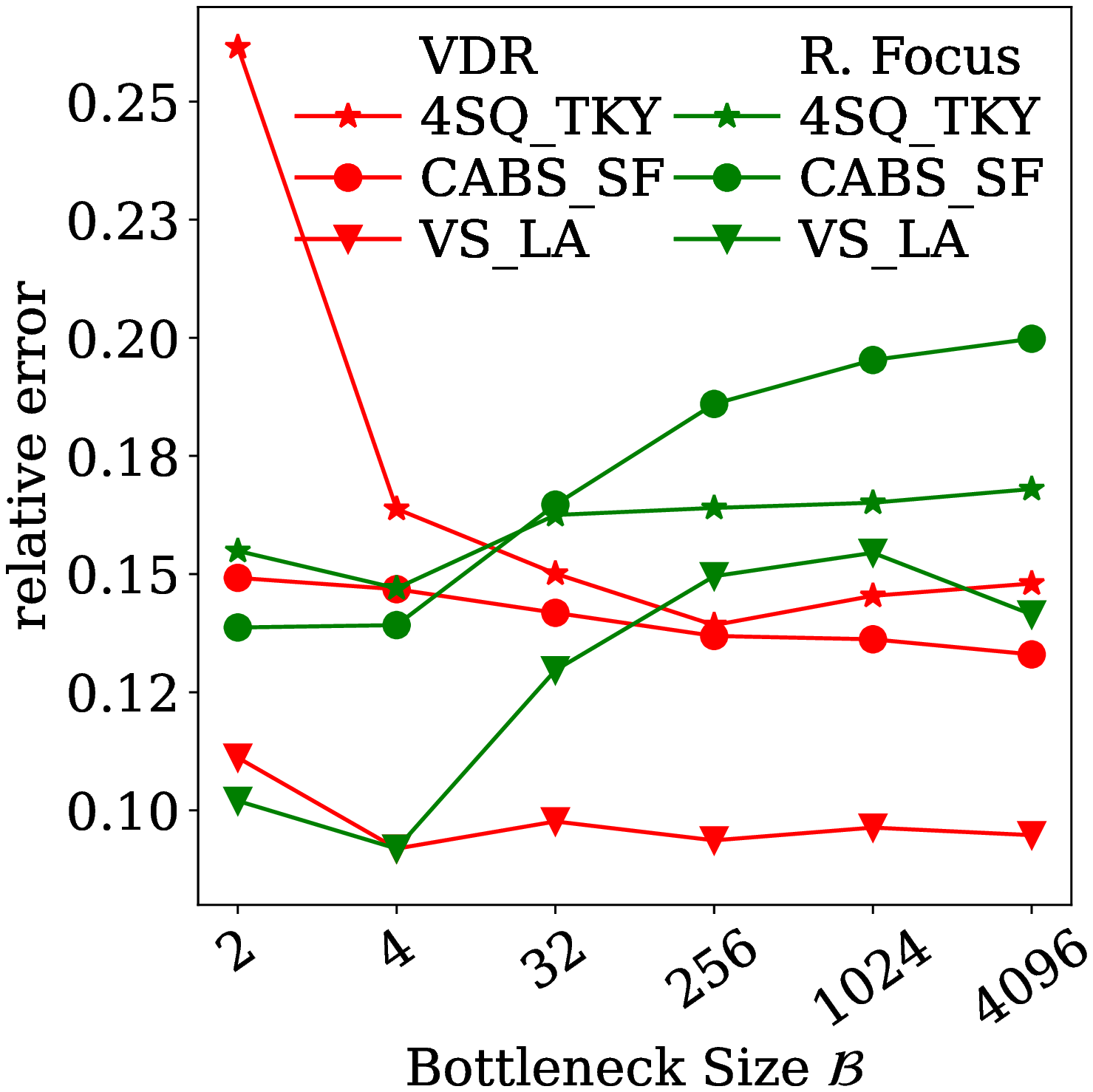}
        \caption{Model regularization analysis}
        \vspace{-2mm}
        \label{fig:regularization_analysis}
    \end{minipage} 
    \hfill
    \begin{minipage}[t]{0.24\textwidth}
        \includegraphics[width=\textwidth]{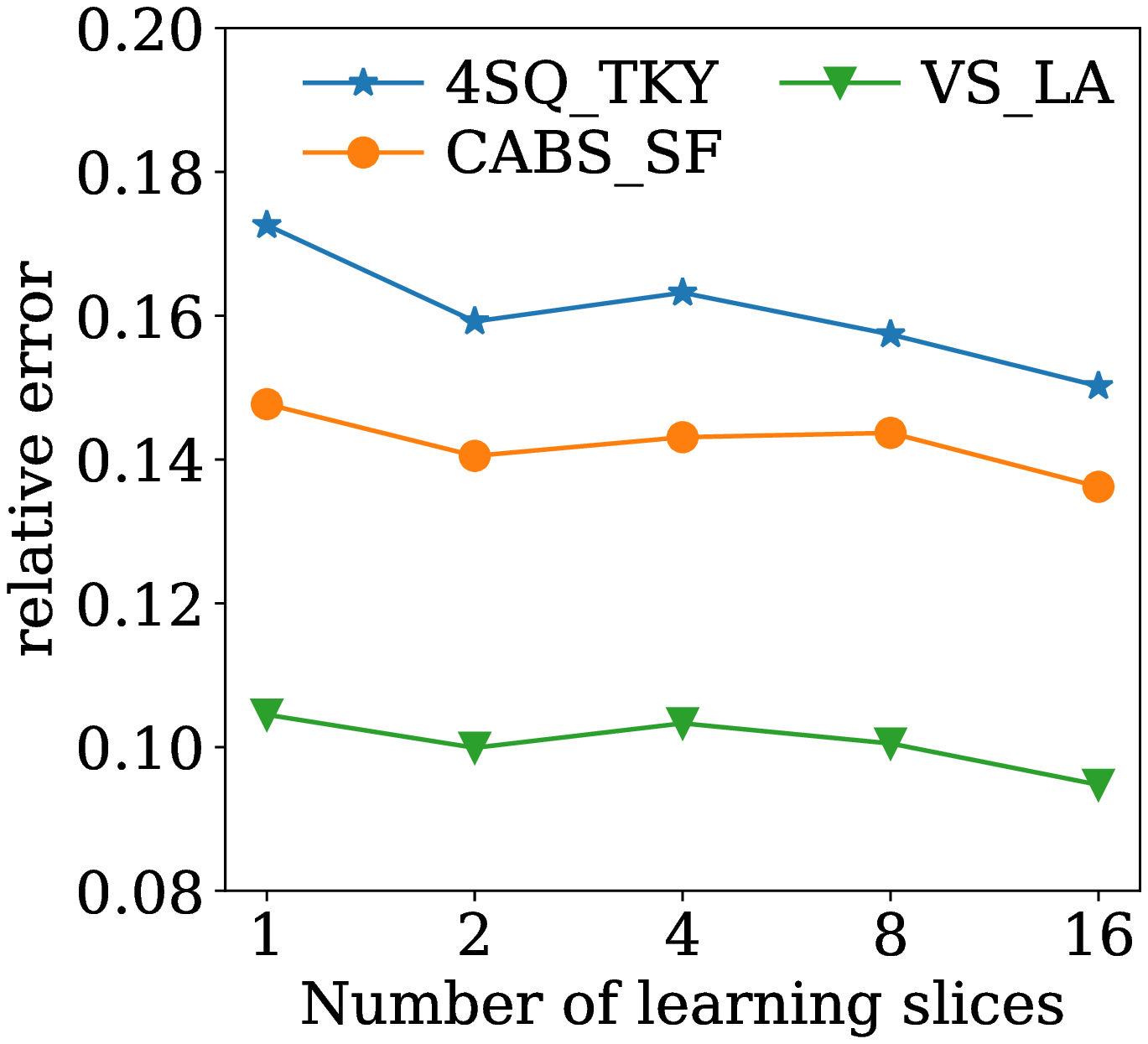}
        \caption{Learning Period Analysis}
        \vspace{-2mm}
        \label{fig:learningperiod_analysis}
    \end{minipage}
        \hfill
    \begin{minipage}[t]{0.24\textwidth}
\includegraphics[width=\textwidth]{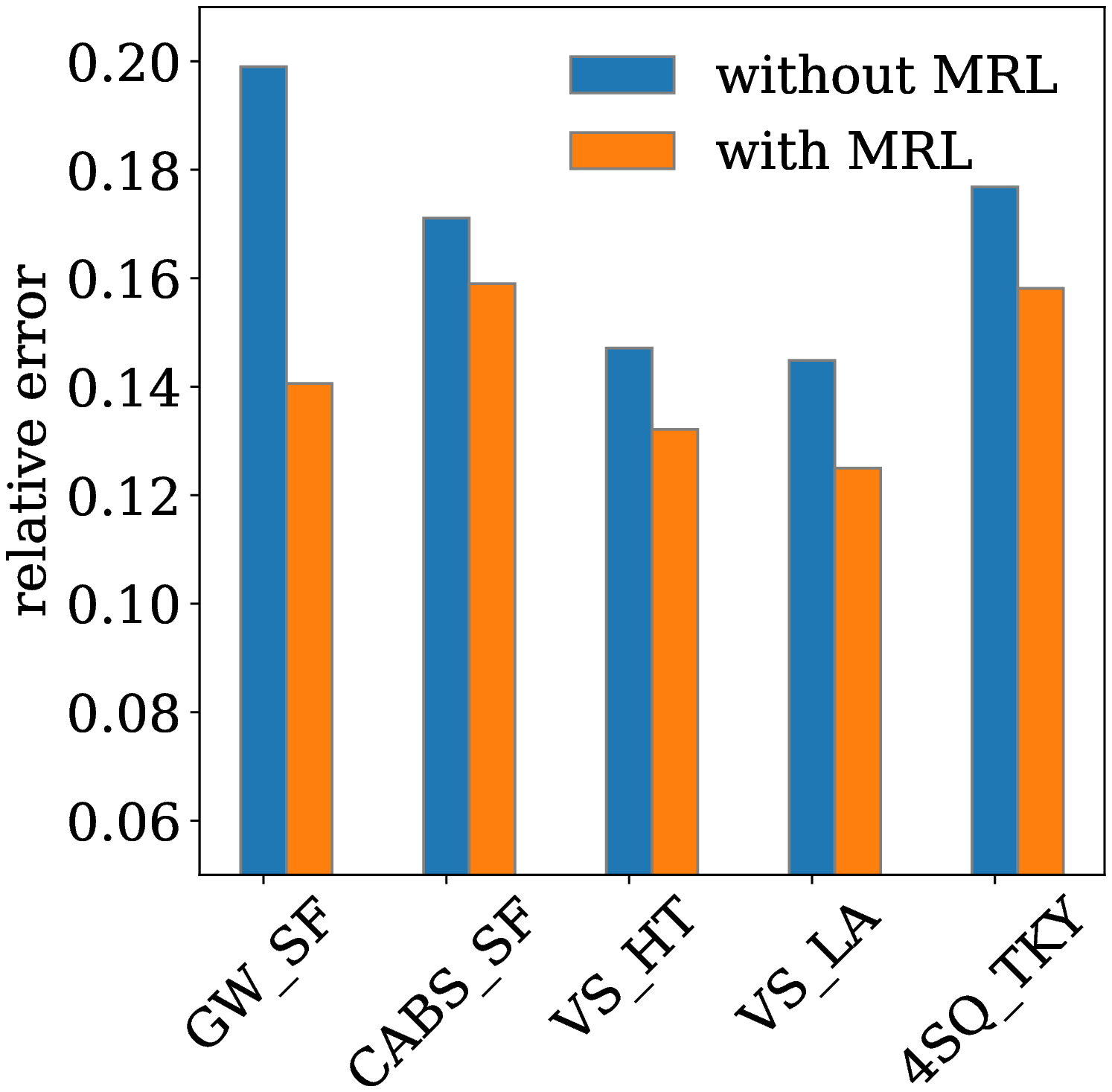}
        \caption{Benefits of Multi Resolution Learning}
        \vspace{-2mm}
        \label{fig:mrl_comparison}
    \end{minipage}
    \hfill
    \begin{minipage}[t]{0.24\textwidth}
\includegraphics[width=\linewidth]{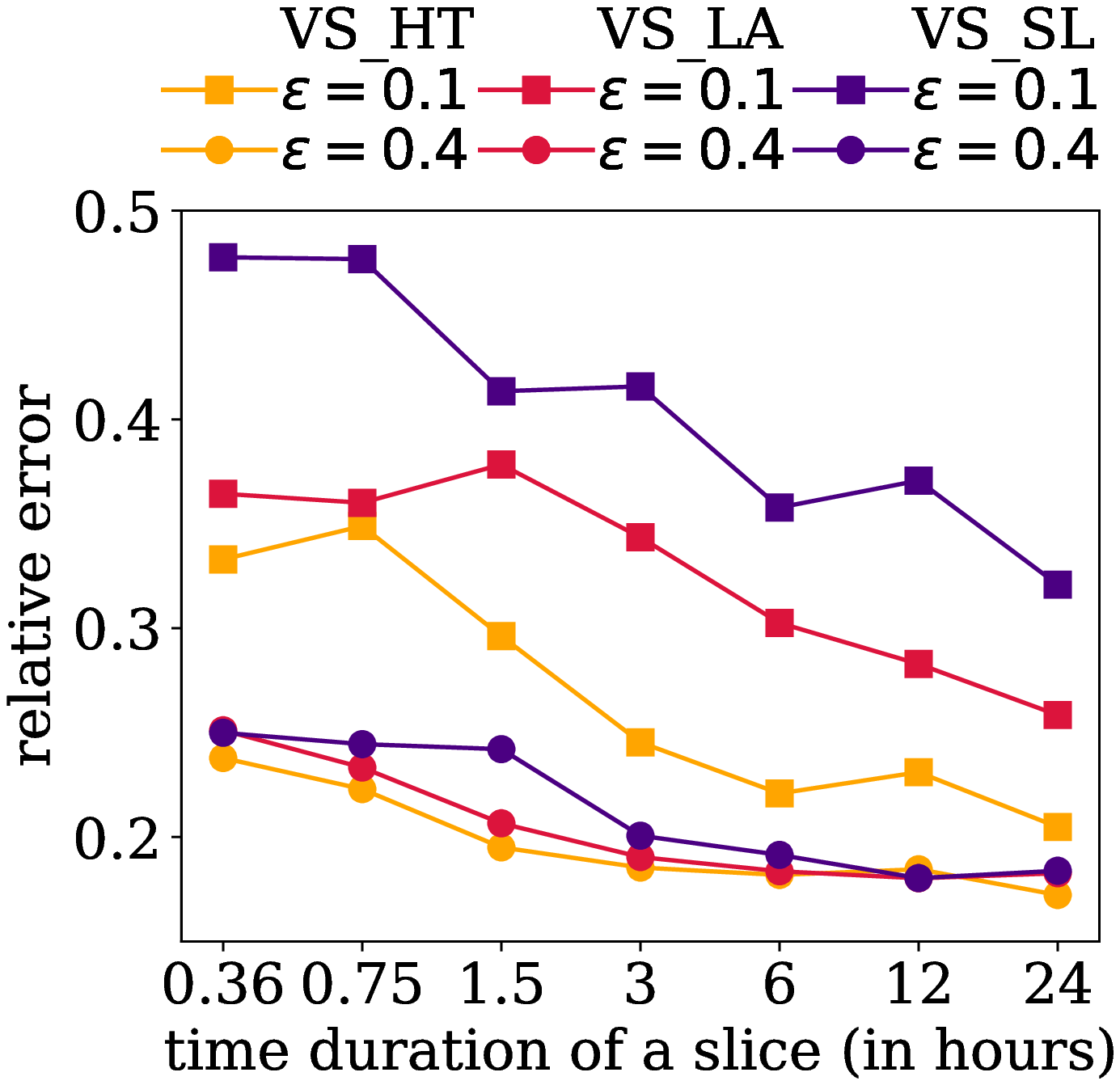}
        \caption{Effect of varying temporal resolution}
        \vspace{-2mm}
    \label{fig:temporal_resolution}
    \end{minipage}
\end{figure*}

\begin{figure}
    \begin{minipage}[t]{0.46\linewidth}
            \includegraphics[width=\textwidth]{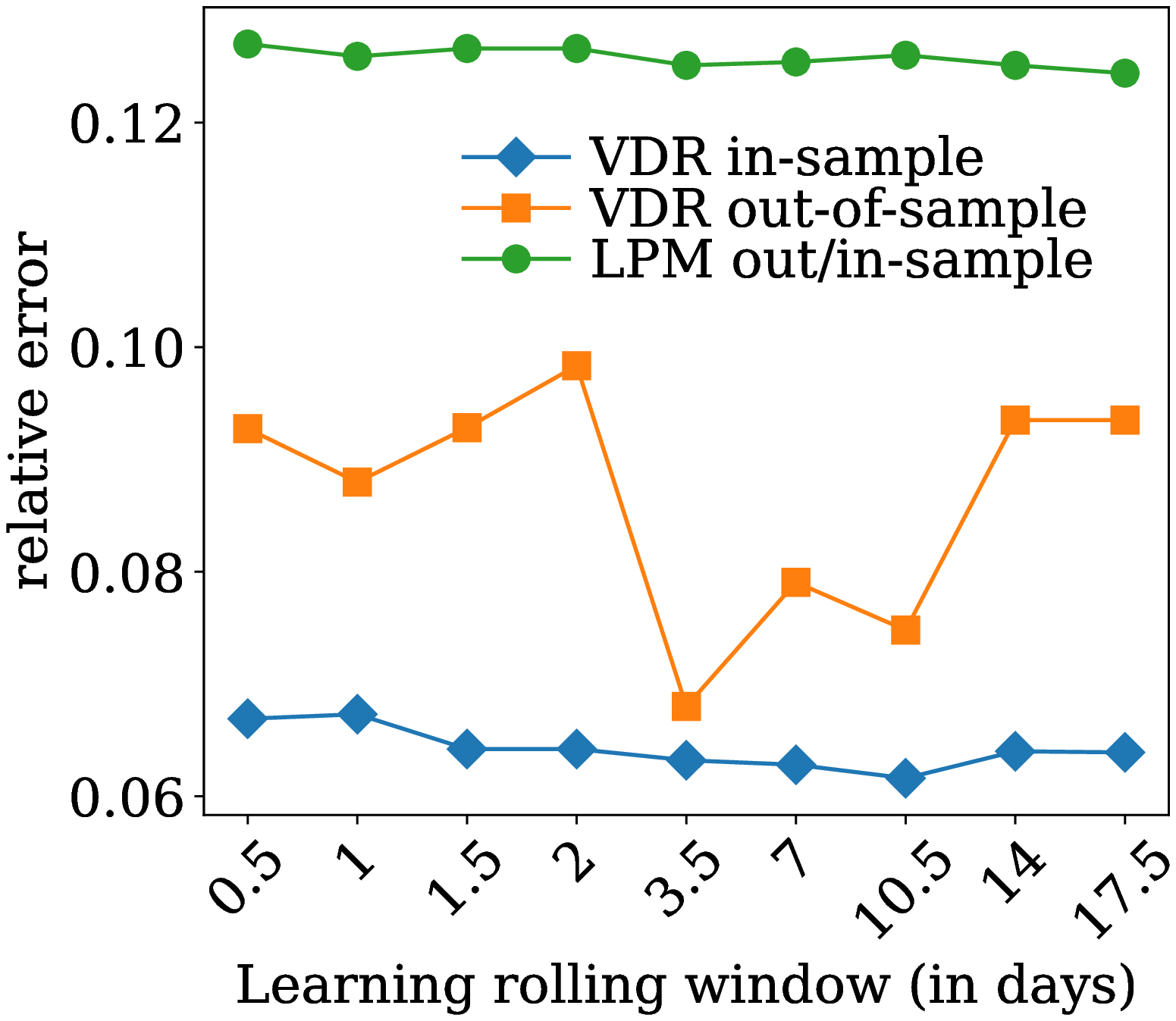}
        \caption{Impact of varying learning period}
        \label{fig:transfer_learning1}
        
    \end{minipage} 
    \hfill
    \begin{minipage}[t]{0.46\linewidth}
            \includegraphics[width=\textwidth]{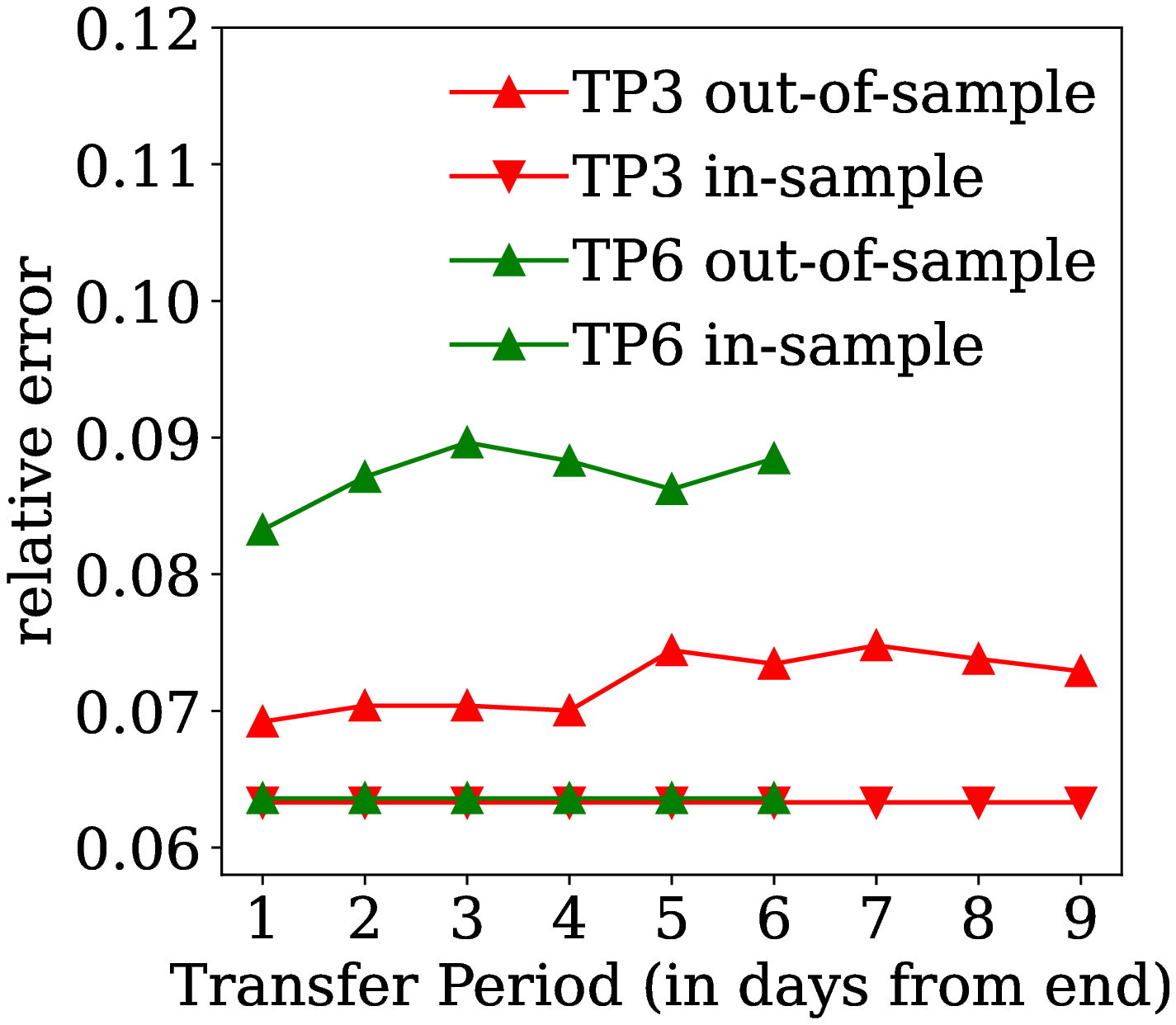}
        \caption{Out-of-sample denoising.}
        \label{fig:transfer_learning2}        
    \end{minipage}
\end{figure}

\subsubsection{Modeling choices\nopunct}\label{subsec:system_modelling}\hfill\\
\noindent\textbf{Effect of model regularization on performance}.
We evaluate the effectiveness of \textit{Variational-AutoEncoders} in denoising DP histograms. Recall that,  by training a lower dimension representation of the data, we wish to learn patterns without overfitting to the noise. In Fig.~\ref{fig:regularization_analysis} we evaluate VDR against `R. focus', a method that simulates an AutoEncoder by over-emphasizing reconstruction loss (i.e., by setting $\alpha$ to a very small value). We consider both public and proprietary datasets while varying the bottleneck size. We notice that a small bottleneck performs poorly due to having limited representation power to learn the input data. When increasing the bottleneck, we see polar effects in the presence and absence of regularization. In the case of `R. Focus', the model quickly overfits to the noise, decreasing accuracy. Whereas if the AutoEncoder is sufficiently regularized, accuracy remains good even for large models due to the learning of generalizable patterns, emphasizing the need for regularizing the encoding space. 


\noindent\textbf{Effect of learning period}.
Fig.~\ref{fig:learningperiod_analysis} shows the accuracy of denoising when 
we train the VAE with a varying number of slices. When the number of learning slices is one, we have in essence a snapshot dataset in 2D. As we add more slices to training, the learning is stabilized and the the learned patterns help achieve better denoising performance in the entire dataset.

\noindent\textbf{Effect of Multi Resolution Learning on accuracy}.
Fig.~\ref{fig:mrl_comparison} shows that across all datasets, by augmenting the training set with coarser granularity histograms, we learn a model that can answer queries more accurately. This is also due to the smoothing effect of the ConvNet, as learned information from one slice helps denoise another within the same dataset.

\noindent\textbf{Effect of Temporal Resolution}.
\revision{We evaluate the accuracy of VDR at several granularities of discretization of the temporal domain. For the 12 day period that each Veraset dataset covers, varying the time duration also varies the number of slices each dataset produces; from as many as 576 (for 22 minutes per slice) to 12 (for 24 hours per slice). Figure \ref{fig:temporal_resolution} shows the accuracy of VDR on all Veraset datasets at privacy budgets of 0.1 and 0.4. Accuracy improves as the temporal resolution becomes coarser. Given that the number of points in the dataset is fixed, there are two compounding effects of increasing the temporal duration per slice. The first increases the number of points in each slice, causing the \textit{relative} error to plummet because the denominator (true answer) grows in value. In addition, the denoising ability of VDR improves with coarser temporal resolution since (1) each slice contains more defined spatial patterns due to longer time aggregation and (2) slices tend to become similar to each other, and the repeating patterns are now easier to summarize for the bottleneck layers of the VAE.}


\subsubsection{Data release over time}
We study the effectiveness of VDR when releasing data over time. Specifically, we measure how often VDR needs to be retrained when new data arrive. For this experiment we utilize the Veraset Houston data for a period of 24 days, with each slice representing a one hour time period.  We consider a training period, $T_b$ to $T_e$, where $T_b$ is the beginning of the training period and $T_e$ is the end, and a testing period that starts at $T_e$ and ends at $T_t$. We refer to the period $T_b$ to $T_e$ as in-sample and $T_e$ to $T_t$ as out-of-sample. We evaluate the performance of the model in two scenarios. In Figure \ref{fig:transfer_learning1}, we test the denoising performance of VDR on an out-of-sample period of 3.5 days (84 slices). VDR in-sample and out-of-sample show the accuracy of VDR on the in-sample and out-of-sample period, respectively. The training data period is varied by moving $T_b$ forward in time but keeping $T_e$ and $T_t$ the same (so training period ranges from 420 to 12 slices). Performance first improves when the training period is up to 3.5 days, as having more data helps the model denoise via better generalizabilty. But as even more data is used in the training, specificity of the patterns is reduced, hence accuracy suffers. In the second setting (Fig.~\ref{fig:transfer_learning2}), we train VDR over slices of `TP' (\textit{Training period}) number of days, and use the trained model to denoise. For in-sample testing the {\em Transfer Period} refers to the accuracy of VDR on the training data itself. We see that the model accuracy is mostly unaffected. For out-of-sample testing \textit{Transfer} period refers to the number of days from $T_e$ to $T_t$. We increase $T_t$ so that transfer period ranges from 1 day to 9 days. We see that performance degrades when denoising out-of-sample periods far from the training periods. We recommend retraining of the VAE model every couple of days.


\subsection{User-level privacy and statistical refinement}\label{sec:exp:ulvl}

\begin{figure}
    \centering
    \begin{minipage}[t]{0.23\textwidth}
        \includegraphics[width=\textwidth]{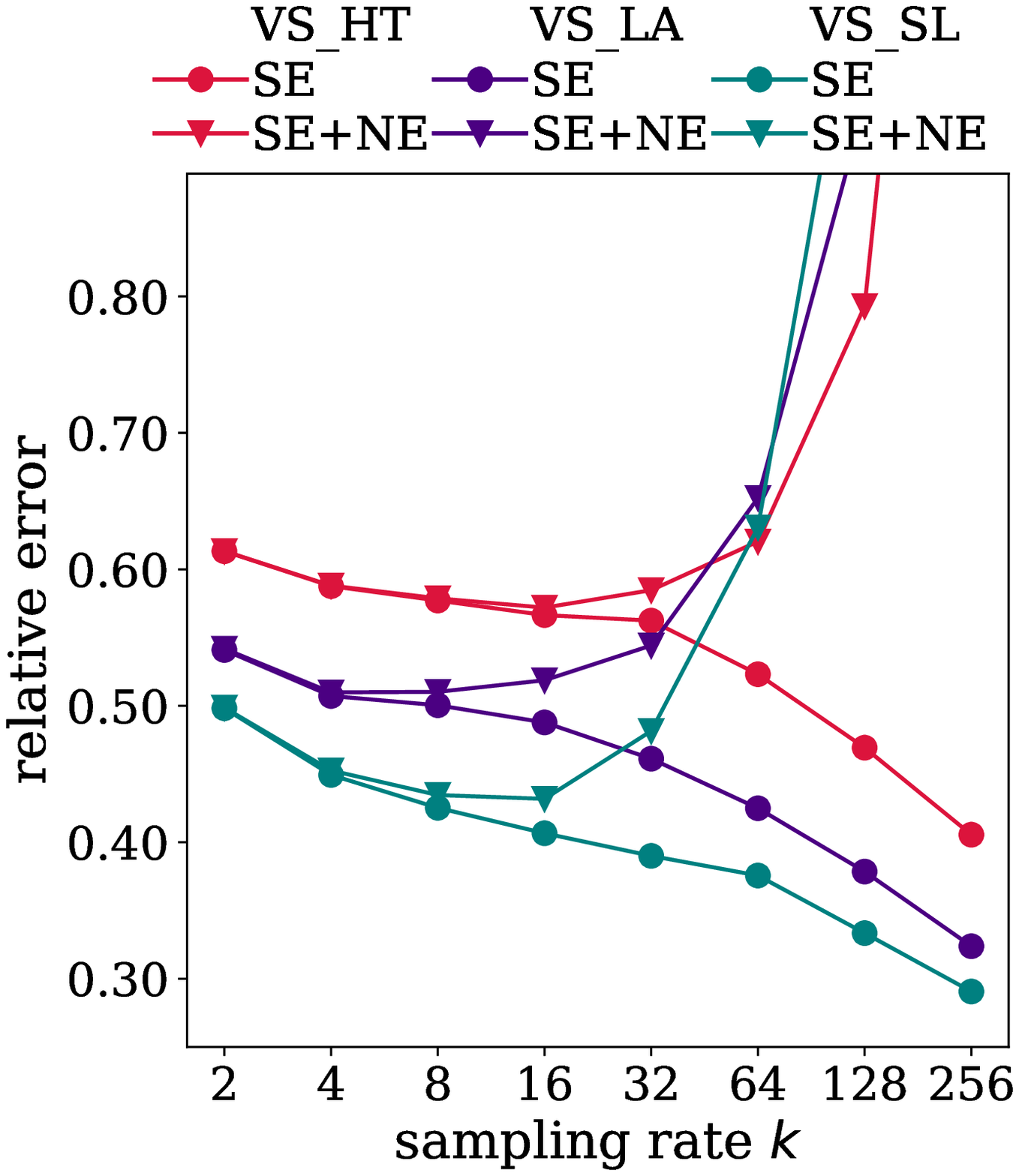}
        \vspace{-2mm}
        \caption{Sampling and Noise Error trends.}
        \label{fig:NEvsSE}
    \end{minipage} 
    \hfill
    \begin{minipage}[t]{0.23\textwidth}
        \includegraphics[width=1\textwidth]{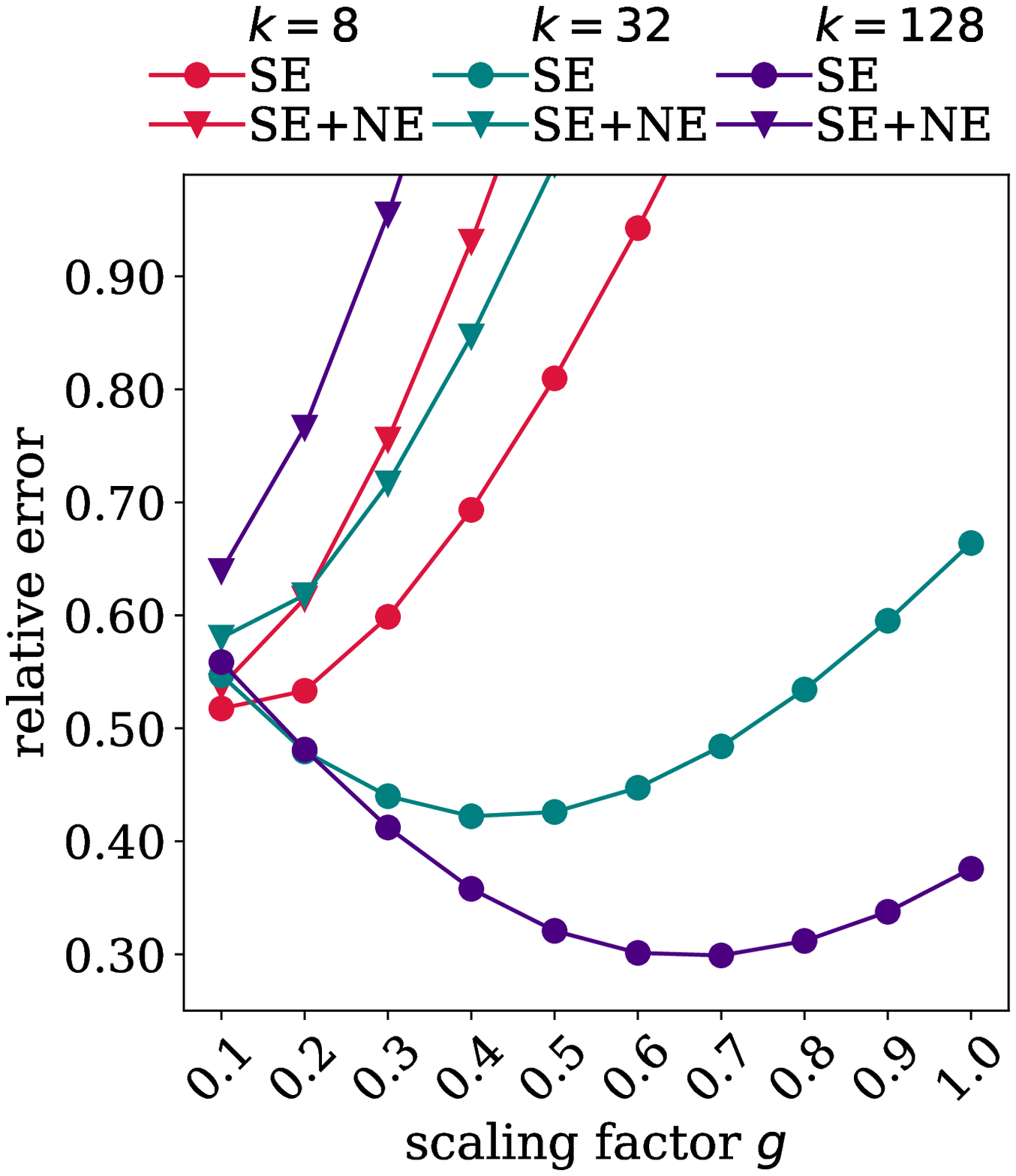}
        \vspace{-2mm}
        \caption{Effect of Scaling factor $g$ on SE and SE+NE }
        \label{fig:debiasing_factor}
    \end{minipage}
    \vspace{-2pt}
\end{figure}
\begin{figure}
    \centering
    \begin{minipage}[t]{0.22\textwidth}
    \includegraphics[width=\textwidth]{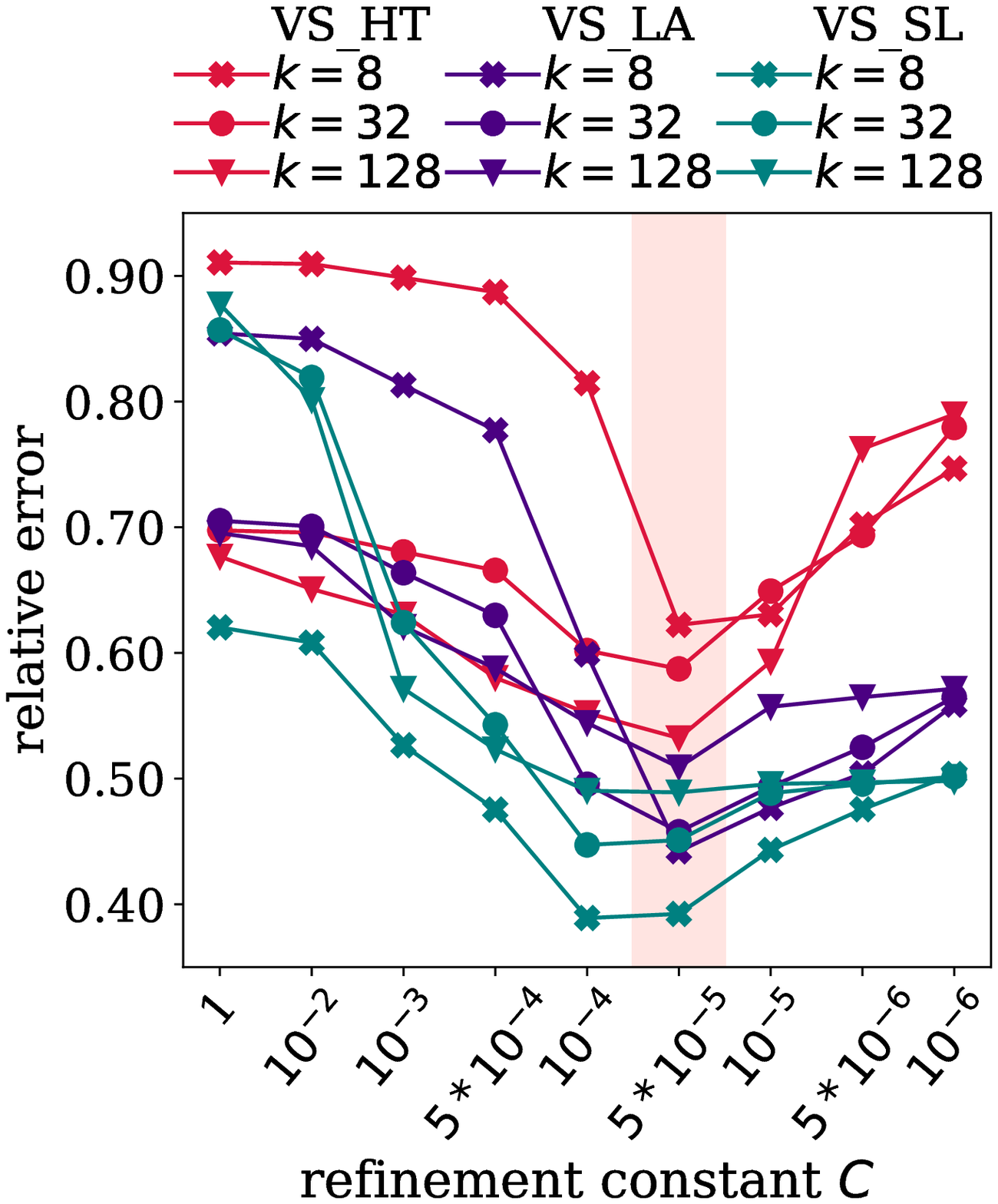}
    \vspace{-2mm}
    \caption{Varying refinement constant $C$ for VDR.}
    \label{fig:vdr_vs_debiasingC}
    \end{minipage}
    \hfill
     \begin{minipage}[t]{0.22\textwidth}
        \includegraphics[width=\linewidth]{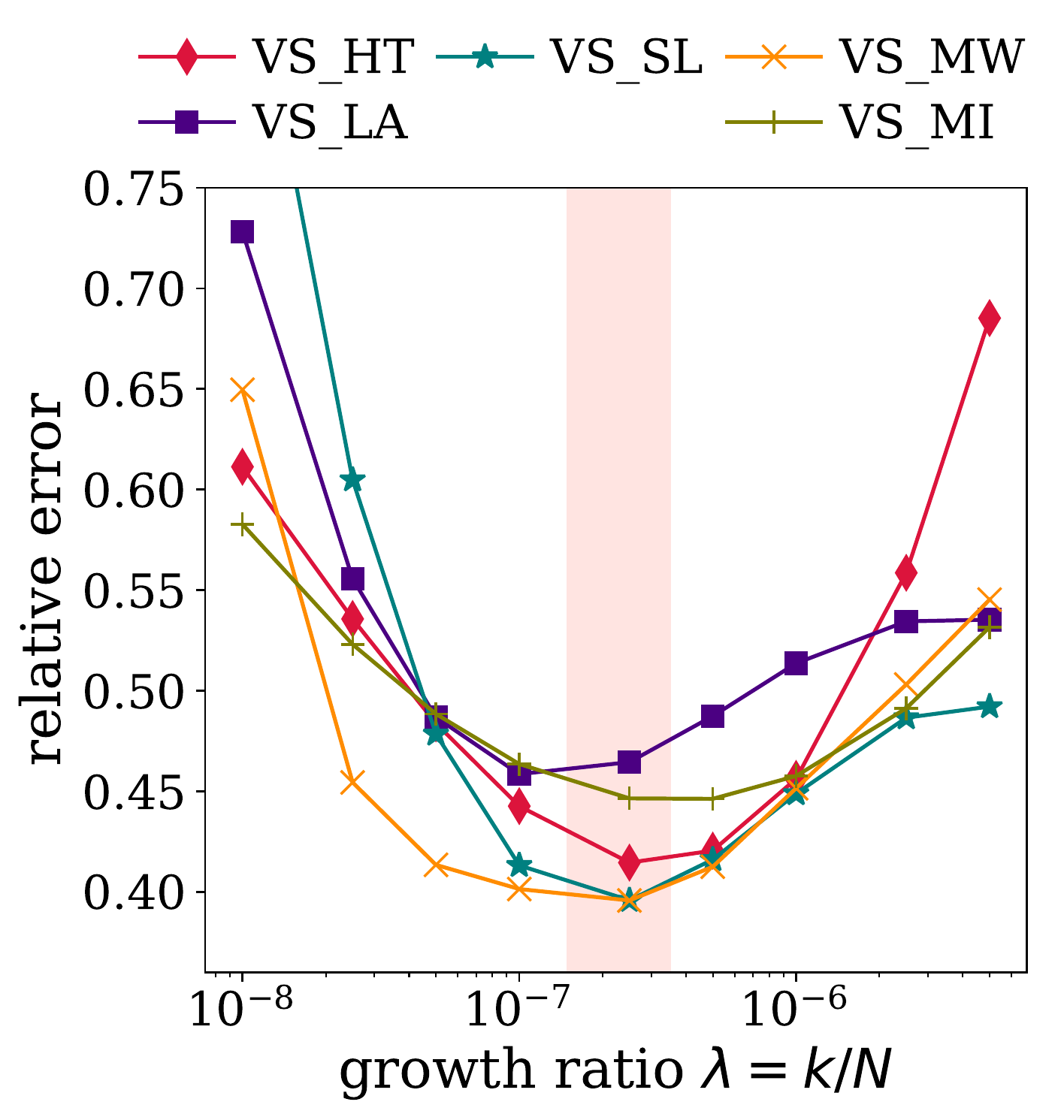}
        \vspace{-2mm}
        \caption{Varying growth ration $\lambda$ for VDR}
        \label{fig:ulvl_growthratio}
    \end{minipage}
    \vspace{-2pt}
\end{figure}

We consider several Veraset data subsets, and set $\varepsilon=6$, analogous to~\cite{aktay2020google, chang2021mobility}. Recall that, in order to release data without consuming excessive privacy budget, we bound the maximum number of contributions of a user to $k$, consequently having access to only the sampled subset $D_s$ to learn our model. Note that in Figs.~\ref{fig:NEvsSE}, \ref{fig:debiasing_factor}, \ref{fig:vdr_vs_debiasingC} and \ref{fig:vdr_debias_performance} the relative error metric is evaluated w.r.t to the true data $D$, while the query computations are performed over $D_s$.

\subsubsection{Challenges for User-Level Privacy}\label{subsec:ulvlchallenges} \revision{We first empirically validate the discussion in Sec.~\ref{sub:collection} and Sec.~\ref{sub:refinement} regarding challenges for user-level privacy without using VDR.}

\noindent\textbf{Bounding user contribution.} \revision{In Fig~ \ref{fig:NEvsSE} we focus on bounding user contributions to achieve good accuracy.} The experiment evaluates the accuracy of answering range count queries  when varying the sampling rate $k$. Sampling error (denoted as SE) measures the error induced purely due to bounding the contribution of each user to $k$. As expected, SE decreases as the sampled subset $D_s$ comes closer to representing the true dataset $D$.

\noindent\textbf{Analysing the effects of brute-force debiasing.} \revision{In Fig~ \ref{fig:debiasing_factor} we show that without VDR, debiasing by scaling the query answers leads to poor accuracy, as discussed in Sec.~\ref{sec:refine:challanges}.} Recall that the answer to the range count query reported on $D_s$ can be scaled according to $N/n$ to potentially debias the result. However, since the data is skewed and with added noise, such a scaling affects the results negatively. We vary the degree of scaling as $g\times N/n$ for values of $g$ from 0.1 to 1. Figure \ref{fig:debiasing_factor} shows that for sampling induced error SE, scaling the answer can be useful. However, after adding DP-compliant noise (plot line SE+NE), upscaling also amplifies the noise in the reported counts and almost always yields poor accuracy. Therefore, to utilize any form of scaling it is important to first denoise the data, as done in VDR.

\subsubsection{VDR System Parameters}\label{sec:exp:ulevel:params}
\revision{Before being able to use VDR to address the above challenges, we need to set the two system parameters: refinement factor $C$ and growth ratio $\lambda$. We empirically validate our guidelines for setting the two parameters here.}

\textbf{Refinement Factor, $C$}. \revision{ In VDR, the denoised histogram is scaled according to the proposed statistical refinement step, in order to offset the effects of sampling at $k$ points per user. The refinement constant $C$ determines, according to Eq.~\eqref{eq:gamma}, the degree of scaling $\gamma$ that is applied to the query answer. For example, at $C=1$, $\gamma$ approaches $N/n$, equivalent to a basic scaling of the query answer. } In Figure \ref{fig:vdr_vs_debiasingC}, we evaluate the accuracy of VDR while varying $C$ at various degrees of sampling $k$. Remarkably, among all settings the lowest error is achieved at $C=5\text{e-5}$, substantiating our claim that a fixed value of $C$ is sufficient to refine answers. \revision{The suitability of a fixed value constant is owing to the fact that human mobility has similar characteristics across cities, as discussed in Sec.~\ref{sec:data_model}. Thus, we recommend to set $C=5\text{e-5}$ across all datasets.}

\textbf{Growth Ratio, $\lambda$}. \revision{Recall that VDR samples $k$ points per user, where $k=\lambda N$ and $\lambda$, the growth ratio, is a system parameter.  In Fig.~\ref{fig:ulvl_growthratio} we vary the growth ratio for several datasets including datasets assumed public for the sake of privacy accounting, VS\_MW (Milwaukee) and VS\_MI (Miami). The results show that across all datasets, similar growth ratio achieves the best accurcy. This confirms that, due to similarity in skewness inherent to location datasets (discussed in Sec.~\ref{sec:data_model}), the same value of $\lambda$ performs well across datasets. Thus, we recommend setting $\lambda = 2.5\text{e-7}$. }

\begin{figure}
\centering
        \includegraphics[width=0.7\linewidth]{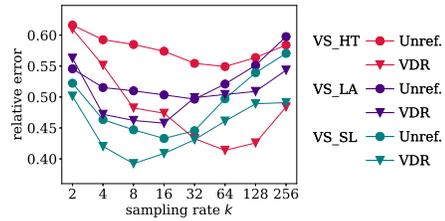}
        \caption{Benefit of VDR refinement }
        \label{fig:vdr_debias_performance}
          \vspace{-2pt}
\end{figure}

\begin{figure*}
\centering
    \begin{minipage}[t]{0.48\textwidth}
\centering
        \includegraphics[height=2.3cm]{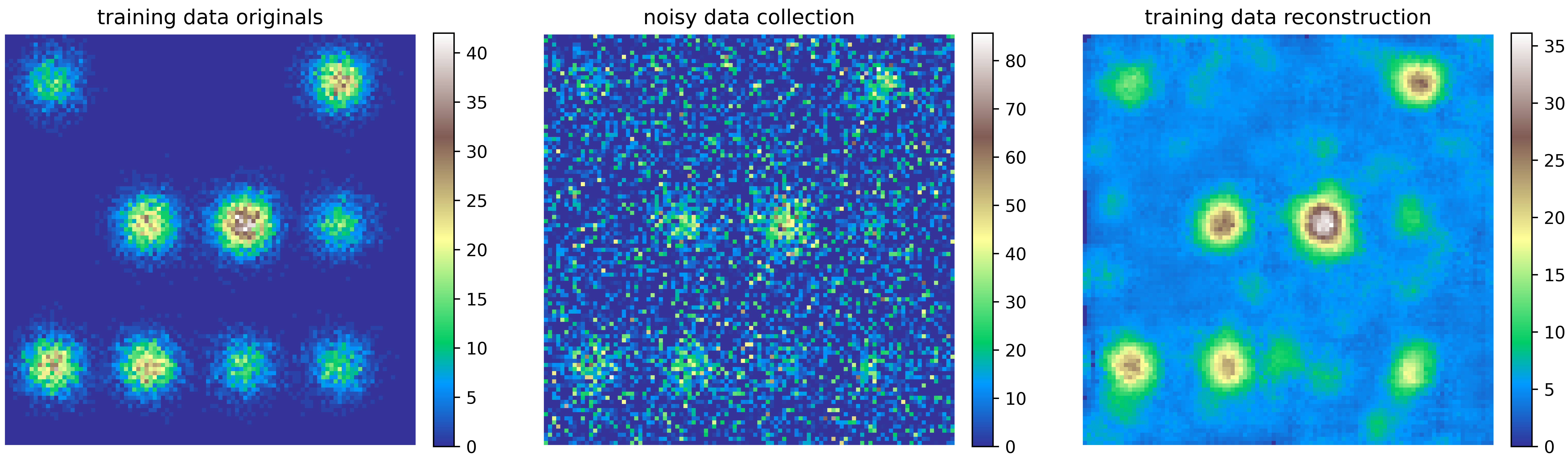}
        \caption{Learning behavior at GMM $\sigma=3$}
        \label{fig:gmm3}
    \end{minipage}
        \hfill\begin{minipage}[t]{0.48\textwidth}
\centering
        \includegraphics[height=2.3cm]{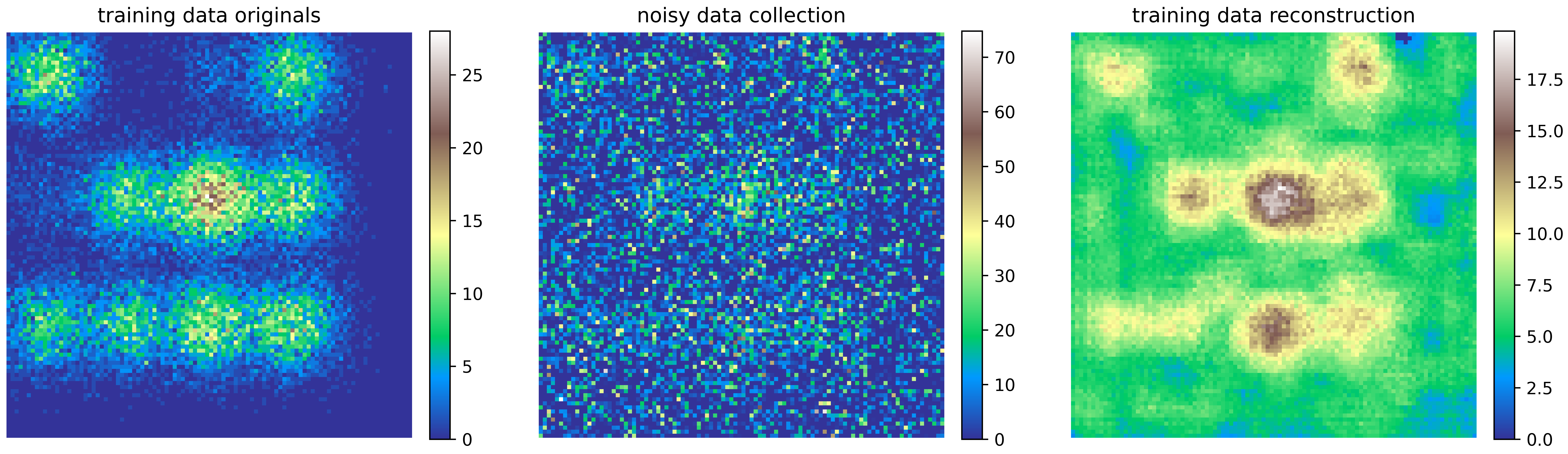}
        \caption{Learning behavior at $\sigma=7$}
        \label{fig:gmm7}
    \end{minipage}
    \begin{minipage}[t]{0.18\textwidth}
    \end{minipage}
    \label{fig:gmm}
    \vspace{-5pt}
\end{figure*}

\begin{figure}
\begin{minipage}[t]{0.2\textwidth}
        \includegraphics[width=\textwidth]{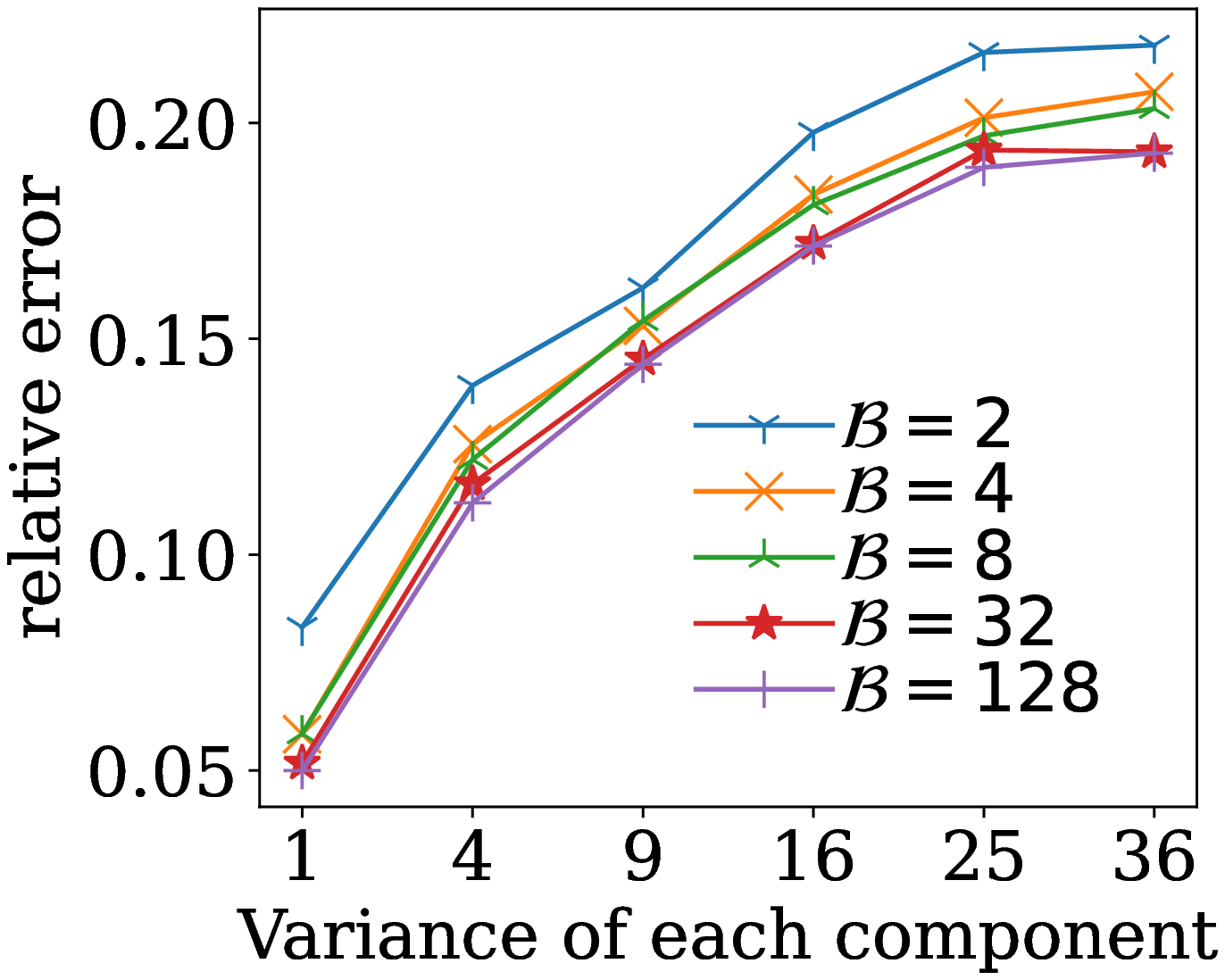}
        \caption{Varying $\sigma$}
        \label{fig:gmm_variance}
        \end{minipage} 
        \hfill
    \begin{minipage}[t]{0.20\textwidth}
        \includegraphics[width=\textwidth]{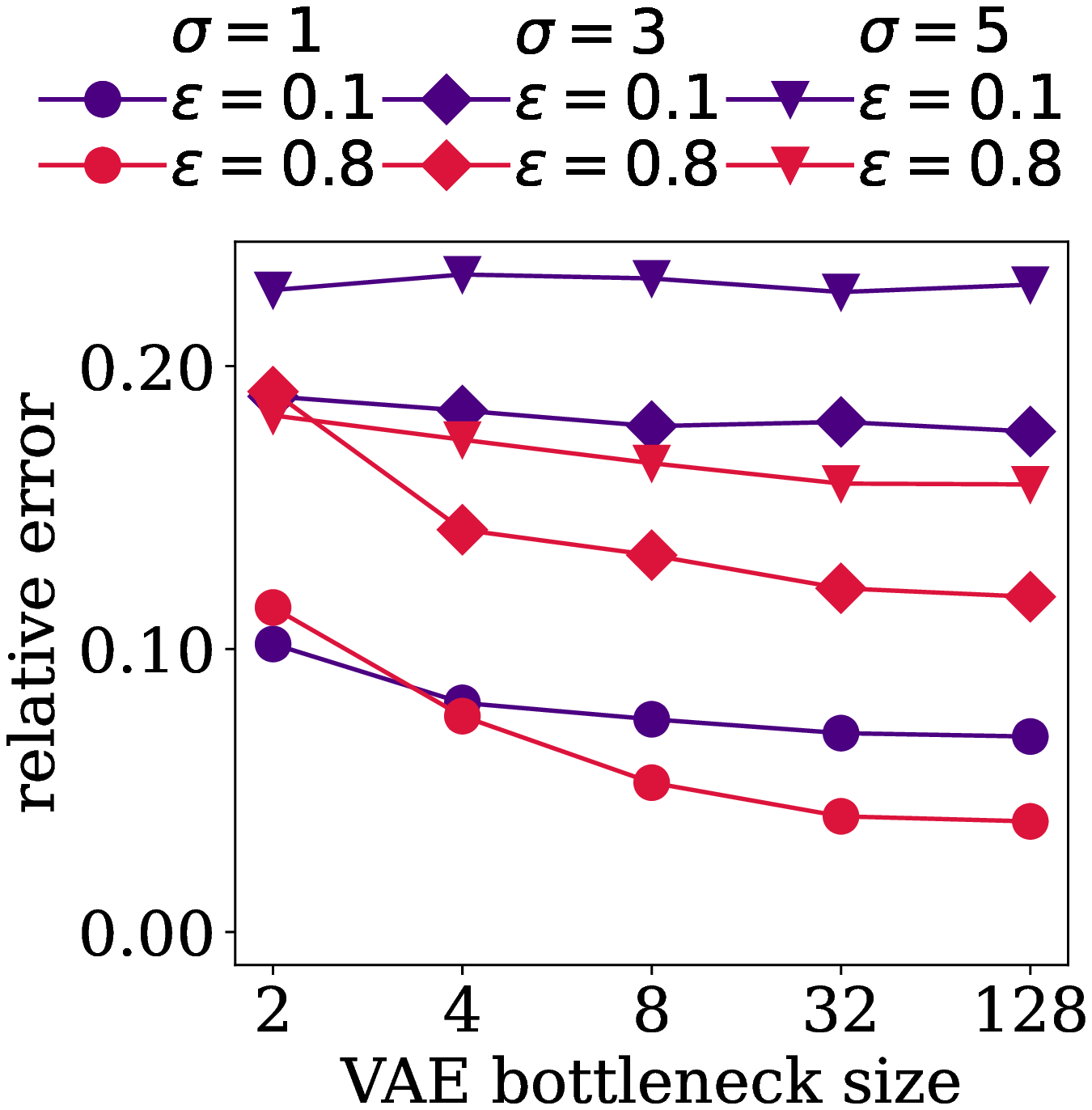}
        \caption{Varying $\mathcal{B}$.}
        \label{fig:gmm_bottleneck}
    \end{minipage}
    \vspace{-2pt}
\end{figure}

\subsubsection{VDR Refinement}  In Fig.~\ref{fig:vdr_debias_performance} we compare VDR with the approach that reports the answer computed on $D_s$ as-is. In all datasets, there is a clear benefit to using statistical refinement, improving the error by up to 40\% in the case of VS\_HT, a high density city. 

 \revision{We conclude with evaluating our heuristic on setting $k$ as  $k=\lambda N$, discussed in Sec.~\ref{sec:system:trade-offs}.  The heuristic, based on the growth ratio $\lambda=2.5\text{e-7}$, empirically determined in Sec.~\ref{sec:exp:ulevel:params},} recommends setting $k= 10$ for VS\_SL, $k=20$ for VS\_LA and $k=53$ for VS\_HT. As we see in Fig.~\ref{fig:vdr_debias_performance}, these values of $k$ achieve close to the best accuracy for their corresponding cities, validating our heuristic for setting $k$ as a constant fraction of data size.



\subsection{Learning Ability on Non-Uniform Datasets}\label{sec:exp:patterns}


\noindent\textbf{Setup}. We synthesize $2M$ points from a Gaussian Mixture Model (GMM) \cite{reynolds2009gaussian} with 50 components positioned at random in the 3D integer lattice $\mathbb{Z}^3$ of size $9$$\times$$9$$\times$$9$. All components are equally weighted and have the covariance matrix $\vec{I}\times\sigma^2$, where $\vec{I}$ is the identity matrix. To control the variation around its mean value, we adjust the parameter $\sigma$. \revision{For clarity of visualization, we partition the synthetic data} into a 3D histogram of 100$\times$100$\times$100 cells. We train and denoise with VDR on the 100 slices. 
We report $\sigma$ in terms of the number of such cells, with a smaller variance implying a data spread tighter around the mean of each GMM component, thus mimicking the skewed data distributions typically present in spatio-temporal location datasets. Fig.~\ref{fig:gmm3} ($\sigma=3$) and Fig.~\ref{fig:gmm7} ($\sigma=7$) visualize a single slice of this dataset with its true values (left), noisy data collections (middle) and denoised reconstructions (right). VDR has a strong ability to recover the underlying patterns of GMMs from even highly distorted observations.
\revision{Moreover, Figure \ref{fig:gmm_variance} ($\varepsilon = 0.2$) shows that as we increase the variance $\sigma^2$ of the GMM components, the model performance suffers, since at a large variance (e.g., Fig.~\ref{fig:gmm7}) data is more uniformly distributed and lacks the spatial patterns typically exhibited in location datasets (such as those depicted in Figure \ref{fig:histogram_overtime}). }
Lastly, we evaluate the effect of varying the bottleneck size of the VAE on the learning ability of VDR. Figure \ref{fig:gmm_bottleneck} shows that, for a given privacy budget, a larger bottleneck is required to capture more skewed datasets. When data are skewed (compare lines for $\varepsilon=0.1$ and $\varepsilon=0.8$ at $\sigma=1$), less DP noise in the data collection step helps further emphasize the data spread, benefiting from having a larger model capacity to learn precisely such patterns. 

\section{Related Work}\label{sec:rel_works}
\noindent\textbf{Private Data Release}.
Longitudinal release of individual location updates increase risk of attack \cite{xiao2010differential, cao2017quantifying}, and requires more stringent privacy settings, e.g., user-level privacy. The work in~\cite{acs2014case} models disjoint regions of the space as separate 1-d time series. However, this limits supported query types, and cannot answer range or hotspot queries. Moreover, the granularity used is very coarse.
PrivBayes~\cite{zhang2017privbayes} is a mechanism that privately learns a Bayesian network over the data, and then returns a matrix used for fitting the parameters of the Bayes net. This can be used to then generate a synthetic dataset which can consistently answer workload queries. Budget allocation is equally split between learning the Bayesian network structure and learning its parameters. Multiplicative-Weights Exponential Mechanism (MWEM) \cite{hardt2012simple} maintains an approximating distribution over the data domain, scaled by the number of records. It updates this distribution by posing a workload of linear queries (e.g., RCQs), finding poorly answered ones, and using the multiplicative update rule to revise its estimates. AHP~\cite{zhang2014towards} seeks to group a histogram’s adjacent bins with close counts to trade for smaller noise. It utilizes LPM, and sets noisy counts below a threshold to zero. The counts are then sorted and clustered using a global clustering scheme to form a partition. SNH \cite{zeighami2022neural} uses a neural database approach to denoise the query answers, by approximating the query answers with a neural network. It requires a workload, which is not available in our setting.


\noindent\textbf{Noise reduction techniques}. 
Most deep-learning based denoising methods \cite{im2017denoising, pang2021recorrupted, lehtinen2018noise2noise} rely on many pairs of clean/noisy images. Denoising autoencoders attempt to learn original data distributions that have been corrupted according to some noise distribution, (e.g., by maximizing the log probability of the clean input, given a noisy input). Recent work in \cite{krull2019noise2void}
trains a model from noisy/noisy image pairs, by extracting noisy versions of the same image repeatedly.  Such a training process is not viable under DP since it would require additional privacy budget for each noisy extraction. Some mild noise from images can also be removed in an unsupervised fashion \cite{zheng2020unsupervised, quan2020self2self}. No approach studied denoising in the presence of DP.

\noindent\textbf{Privacy preserving machine learning}. 
A learned model may leak information about the data it was trained on \cite{shokri2017membership,hitaj2017deep}. Application of DP to empirical risk minimization \cite{chaudhuri2011differentially, kifer2012private} and deep neural networks \cite{sealfon2021efficiently, abadi2016deep} has been recently explored. Existing approaches add noise to the output of the trained model \cite{wu2017bolt}, add a random regularization term to the objective function ~\cite{chaudhuri2011differentially, kifer2012private}, or add noise to the gradient of the loss function during training~\cite{abadi2016deep}.
Our approach sanitizes the training data {\em before} learning. Furthermore, the work of \cite{abadi2016deep} achieves ($\varepsilon, \delta$)-DP~\cite{nissim2007smooth, erlingsson2014rappor, censusgovKDD18}, a weaker privacy guarantee.

\vspace{-5pt}
\section{Conclusion}
We proposed a technique for accurate DP-compliant release of spatio-temporal histograms that uses a combination of sampling to reduce sensitivity, VAE-based learning to counter the effect of DP-added noise, and statistical estimators to offset the effect of sampling. The resulting approach captures well spatio-temporal data patterns, and significantly outperforms existing approaches. In future work, we plan to extend our work by creating DP-compliant synthetic datasets based on spatio-temporal histograms. This is more challenging, since it needs to take into account any type of downstream processing that may be performed. One direction to achieve this goal is to sample from the compressed latent space conditioned on the time-of-day, and train a conditional image generation model such as PixelCNN~\cite{van2016conditional} over the latent pixel values.

\begin{acks}
This research has been funded in part by NIH grant R01LM014026, NSF grants IIS-1910950, CNS-2125530 and IIS-2128661, and an unrestricted cash gift from Microsoft Research. Any opinions, findings, conclusions or recommendations expressed in this material are those of the author(s) and do not necessarily reflect the views of any of the sponsors such as the NSF.
\end{acks}

\bibliographystyle{ACM-Reference-Format}
\balance
\bibliography{references}

\appendix
\begin{figure*}[ht!]
    \centering
    \begin{minipage}[t]{\textwidth}
        \includegraphics[width=\textwidth]{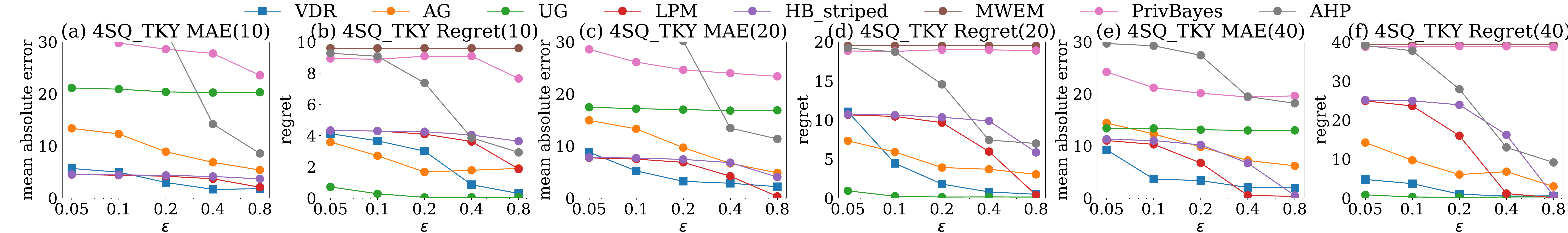}
    \caption{Impact of privacy budget on the hotspot query for the 4SQ Tokyo dataset at various thresholds.}
    \label{fig:hotspot_tky}
    \end{minipage}    
\end{figure*}
\begin{figure*}[btp]
    \begin{minipage}[t]{0.6\textwidth}
        \includegraphics[height=2.5cm]{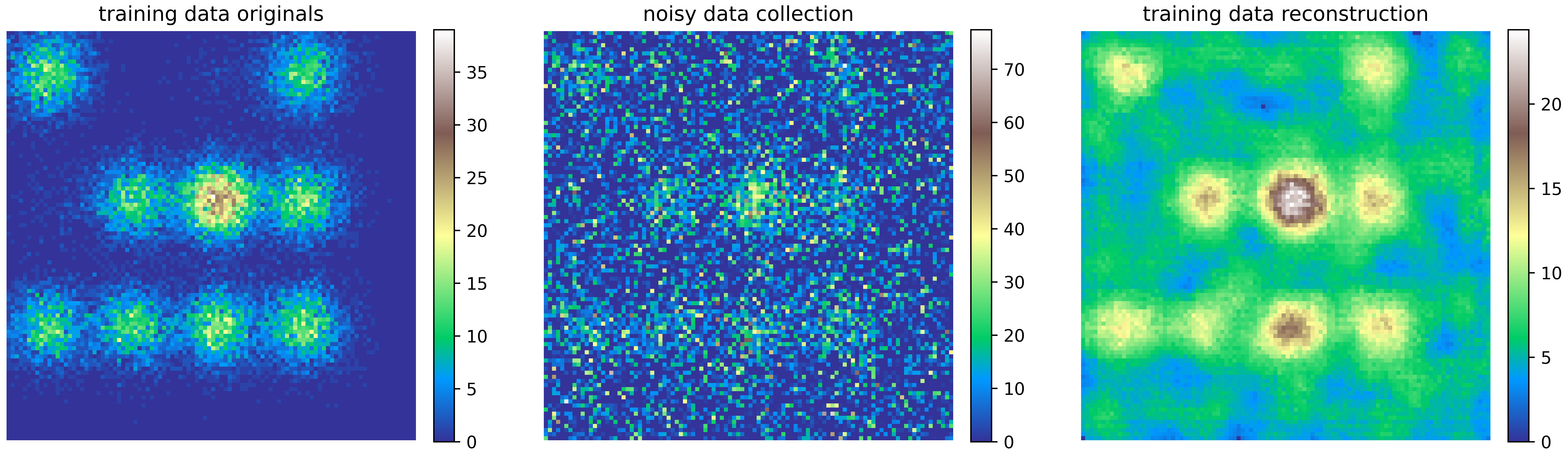}
        \caption{Learning behavior at GMM $\sigma=5$}
        \label{fig:gmm5}
    \end{minipage} 
    \begin{minipage}[t]{0.6\textwidth}
        \includegraphics[height=2.5cm]{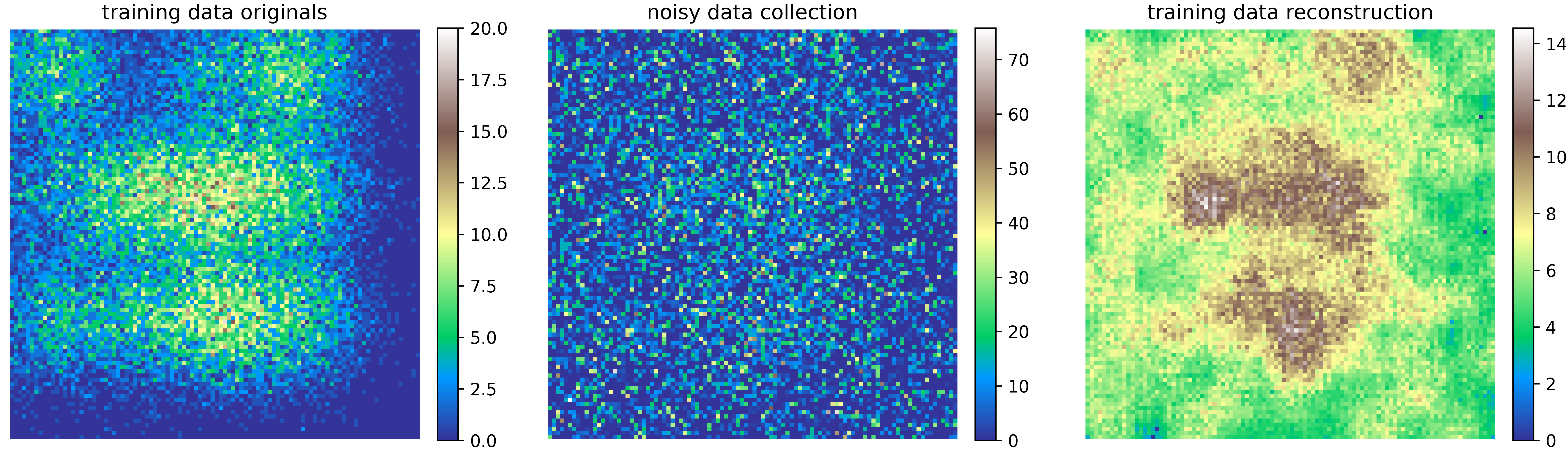}
        \caption{Learning behavior at GMM $\sigma=9$}
        \label{fig:gmm9}
    \end{minipage} 
\end{figure*}

\section{DP Proof}\label{appx:proof_security}
In our discussion, we use the size of the datasets $N$ and sampled sets $n$, which we assume are publicly available and, if not, an estimate can be obtained by spending negligible privacy budget. We also use the well known property of DP: the \textit{post-processing property} of differential privacy~\cite{dwork2014algorithmic} states that given any arbitrary function $h$ and an $\varepsilon$-DP mechanism $\mathscr{M}$, the mechanism $h(\mathscr{M})$ is $\varepsilon$-DP.

\textbf{Proof. }Alg.~\ref{alg:final_alg} shows our proposed end-to-end algorithm. We rewrite the algorithm $\mathscr{M}$ as a composition of mechanisms $\mathscr{M}_s$ and $\mathscr{M}_v$ for the purposes of privacy accounting. $M_s$ is a two step process. The first bounds the number of each user's points to at most $k$. The second step partitions the domain into a 3-d histogram and for each disjoint cell, queries the sampled dataset for the number of points in its extents, and adds noise sampled from the Laplace distribution of scale $\Delta_f/\varepsilon$ to this answer. In the mechanism $\mathscr{M}_v$, VDR takes as input the sanitized 3-d histogram, denoises it, and then, scales the counts according to a statistical refinement heuristic.

$\mathscr{M}_s$ prunes each user's data points to at most $k$, where $k$ is determined in a data-independent manner (as a constant fraction of $N$, see \ref{sec:exp:ulvl} for details). The data domain is partitioned into a 3-d histogram $H$ of granularity $\mathbb{M}$$\times$$\mathbb{M}$$\times$$\mathbb{T}$. For each cell $c\in H$ we write as $f_c$ the query asking the count of points in the cell $c$. We consider these queries as a single vector $\mathbf{Q}$, and pose this query set to be answered over the dataset. Therefore, the $L_1$ sensitivity of $\mathbf{Q}$ is $k$ (i.e., $\Delta_f=k$) and hence the mechanism adding Laplace noise scaled to $k/\varepsilon$ to each unit of $Q$ enjoys $\varepsilon$-DP (Theorem 1 of \cite{kellaris2013practical}). 

Next the mechanism $M_v$ trains the VAE on the sanitized 3-d histogram, and refines the output results \textit{after} the histogram has been sanitized. Therefore the transformation $M_v$ is applied post-sanitization, and due to the post processing property of DP does not consume any privacy budget. Moreover, the statistical refinement step determines scaling factor $\gamma$, according to a constant value of $C$, without running any computation on the private dataset. To conclude, the mechanism $\mathscr{M}$, which is a composition of $\mathscr{M}_s$, $\mathscr{M}_v$ is $\varepsilon$-differentially private.
\hfill$\square$

\begin{table}[ht]
\caption{Curve fit parameters for power-law function  $f(x) = a \times x^b$ on histograms in Figures \ref{fig:datasets_power_law} and \ref{fig:datasets_power_law_vs}}\label{tbl:table_power_law}
\begin{tabular}{@{}lcc@{}}
\toprule
\textbf{Dataset}       & Scalar $a$ & Exponent $b$ \\ \midrule
Houston (VS\_HT)     & 5.16        & -2.69         \\
Los Angeles (VS\_LA) & 4.35        & -2.33         \\
Salt Lake (VS\_SL)  & 4.41        & -2.45    \\
Milwaukee (VS\_MW)   & 5.04        & -2.61     \\ 
Gowalla SF (GW\_SF)   & 2.67        & -1.62     \\ 
Gowalla NY (GW\_NY)   & 2.54        & -1.75    \\ 
Foursquare Tokyo (4SQ\_TKY)   & 3.67        & -1.77    \\ \bottomrule   
\end{tabular}
\end{table}

\section{Additional Experiments}\label{appx:exp}
\subsection{Additional Visualizations for GMM}
Figure \ref{fig:gmm5} and figure \ref{fig:gmm9} visualize additional settings for GMM components.
\subsection{HostSpot Query evaluation}
Figure \ref{fig:hotspot_tky} reports the Mean Absolute Error (MAE) and Regret metrics of the hotspot query for the Foursquare Tokyo dataset at density thresholds of 10, 20 and 40.
\subsection{Dataset Charactaristics}\label{appx:data_charactaristics}
\begin{figure*}[btp]
    \begin{minipage}[t]{0.24\textwidth}
        \includegraphics[width=\textwidth]{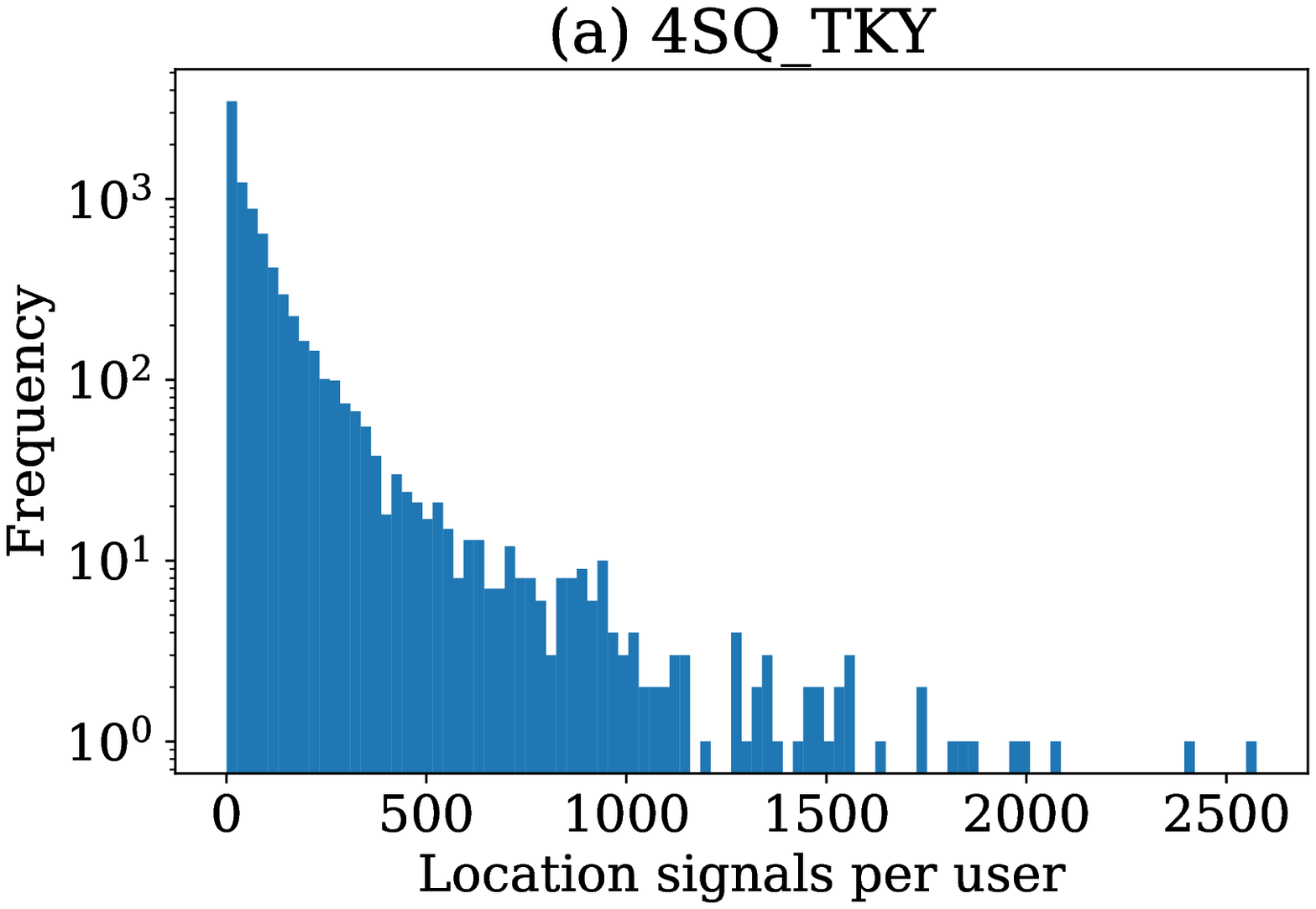}
    \end{minipage} 
    \begin{minipage}[t]{0.24\textwidth}
        \includegraphics[width=\textwidth]{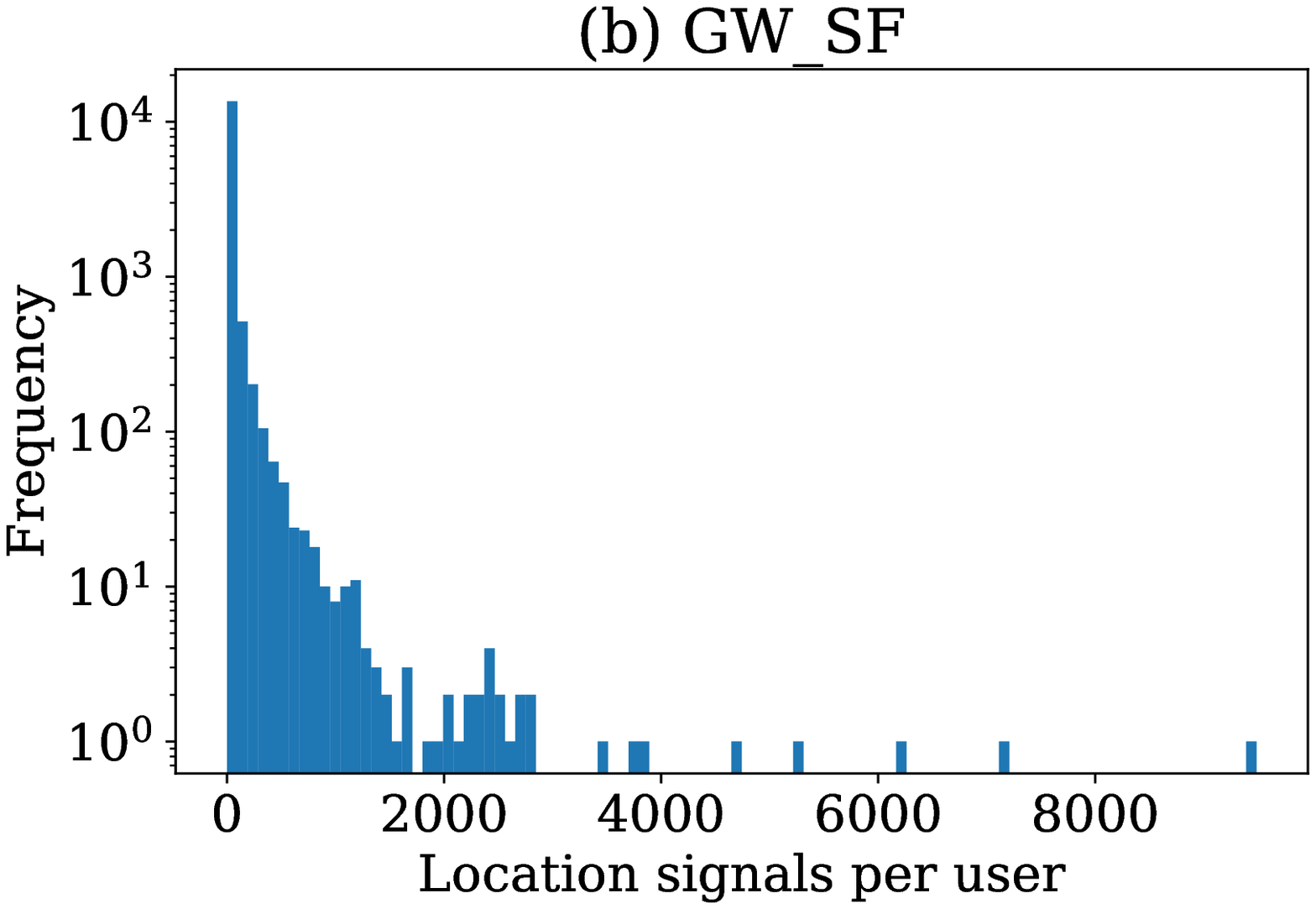}
    \end{minipage} 
    \begin{minipage}[t]{0.24\textwidth}
        \includegraphics[width=\textwidth]{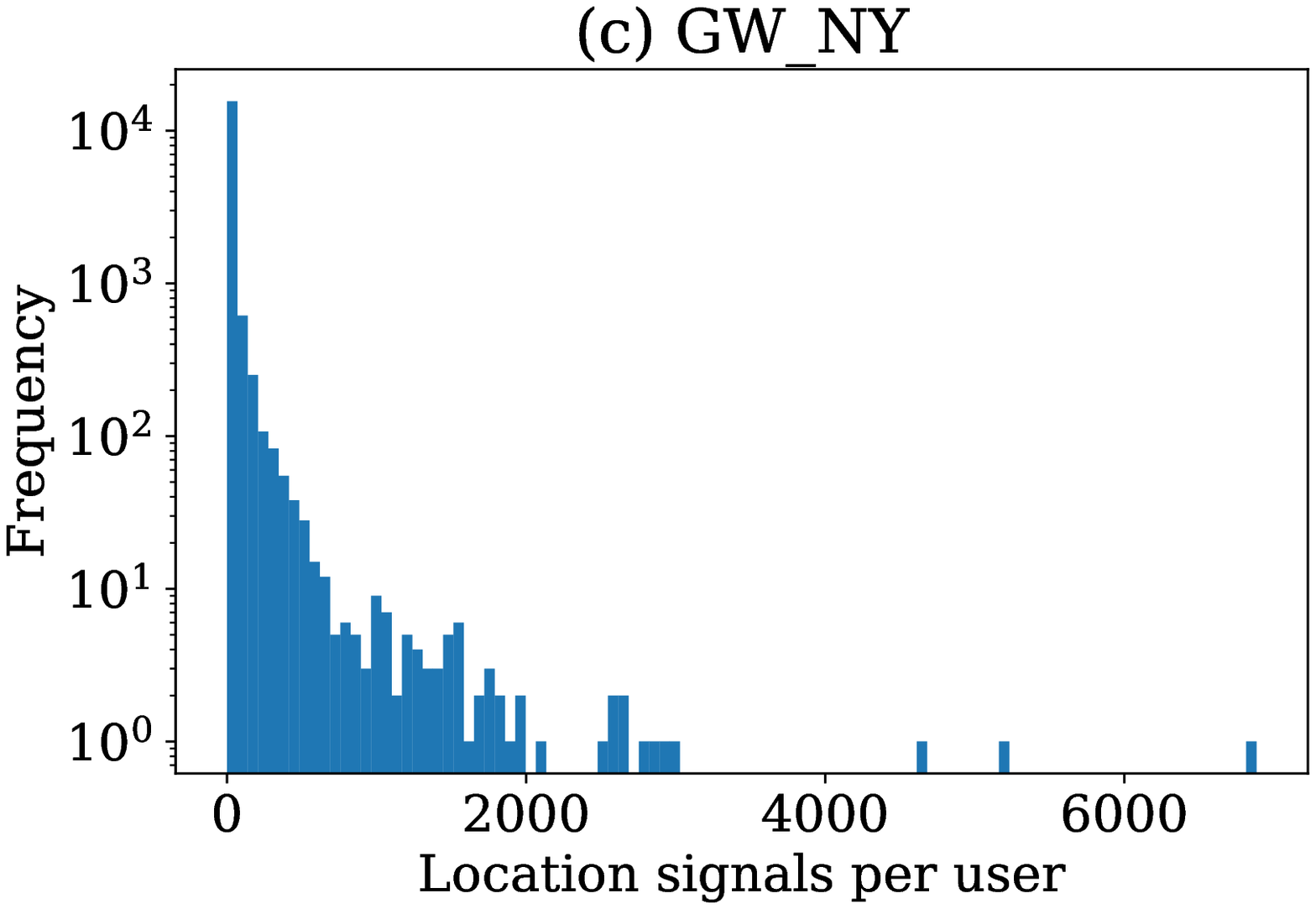}
    \end{minipage} 
    \caption{Histogram of location signals per user in public datasets}
    \label{fig:datasets_power_law}
\end{figure*}

\begin{figure*}[btp]
    \begin{minipage}[t]{0.23\textwidth}
        \includegraphics[width=\textwidth]{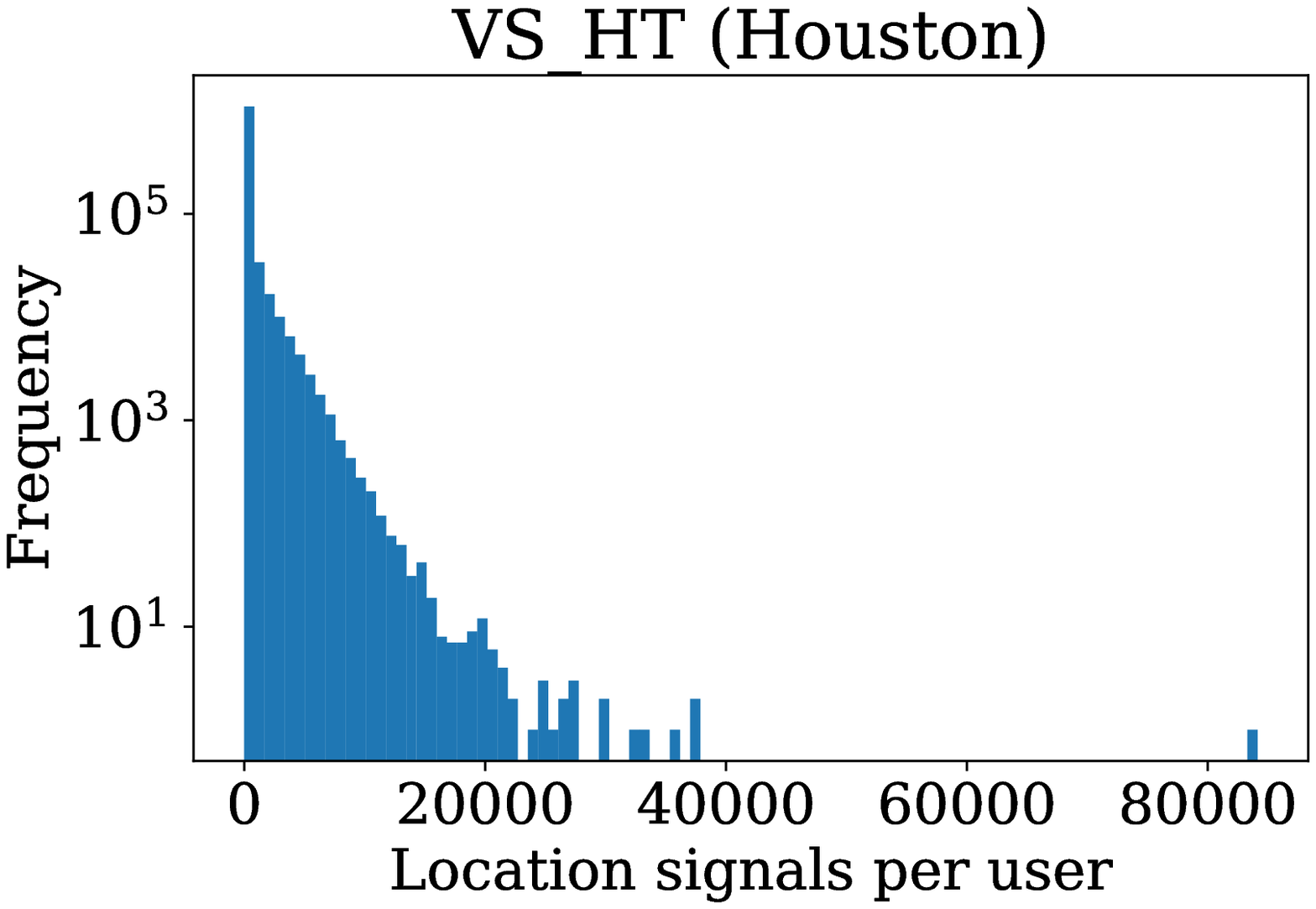}
    \end{minipage} 
    \begin{minipage}[t]{0.23\textwidth}
        \includegraphics[width=\textwidth]{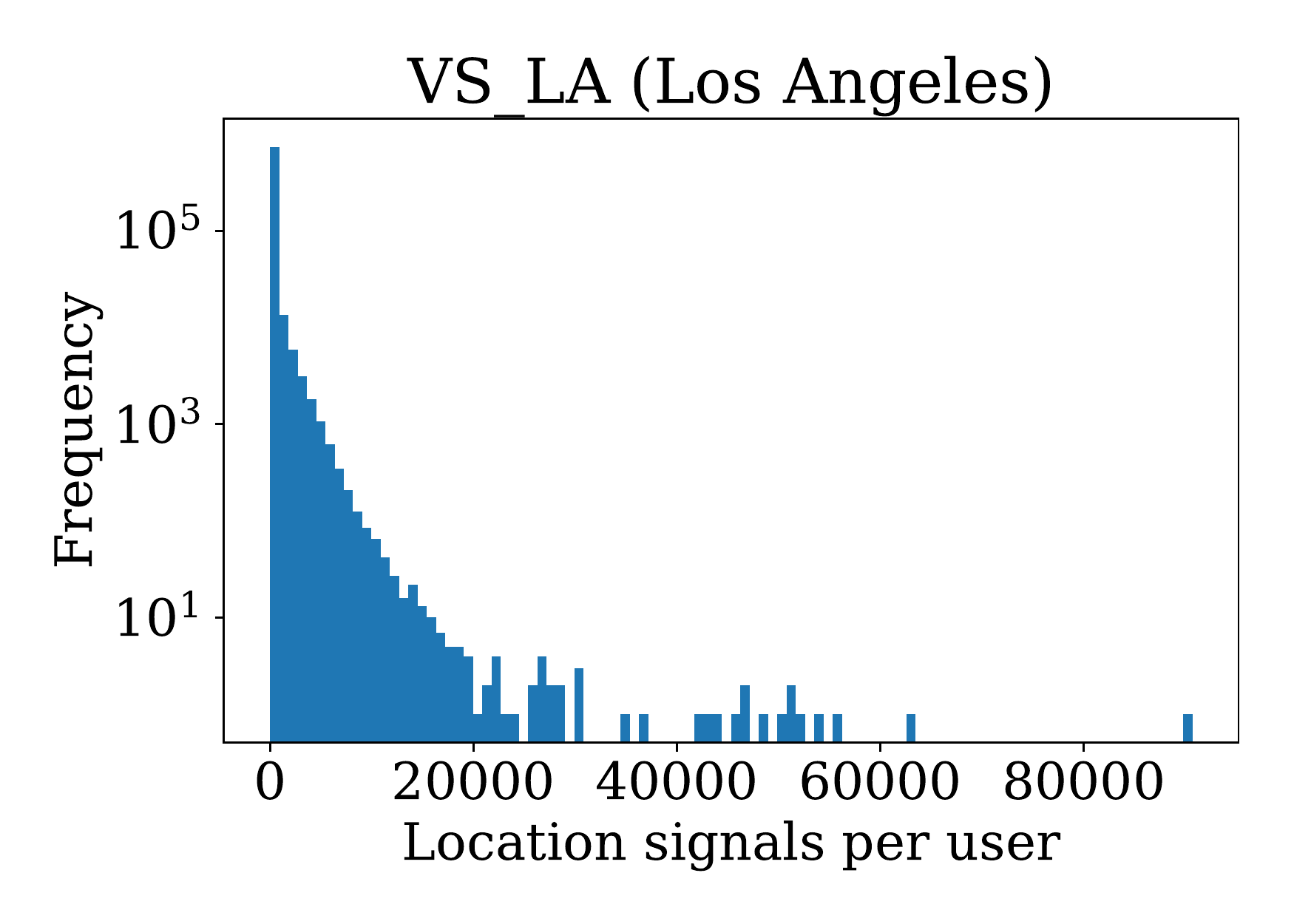}
    \end{minipage}
    \begin{minipage}[t]{0.23\textwidth}
        \includegraphics[width=\textwidth]{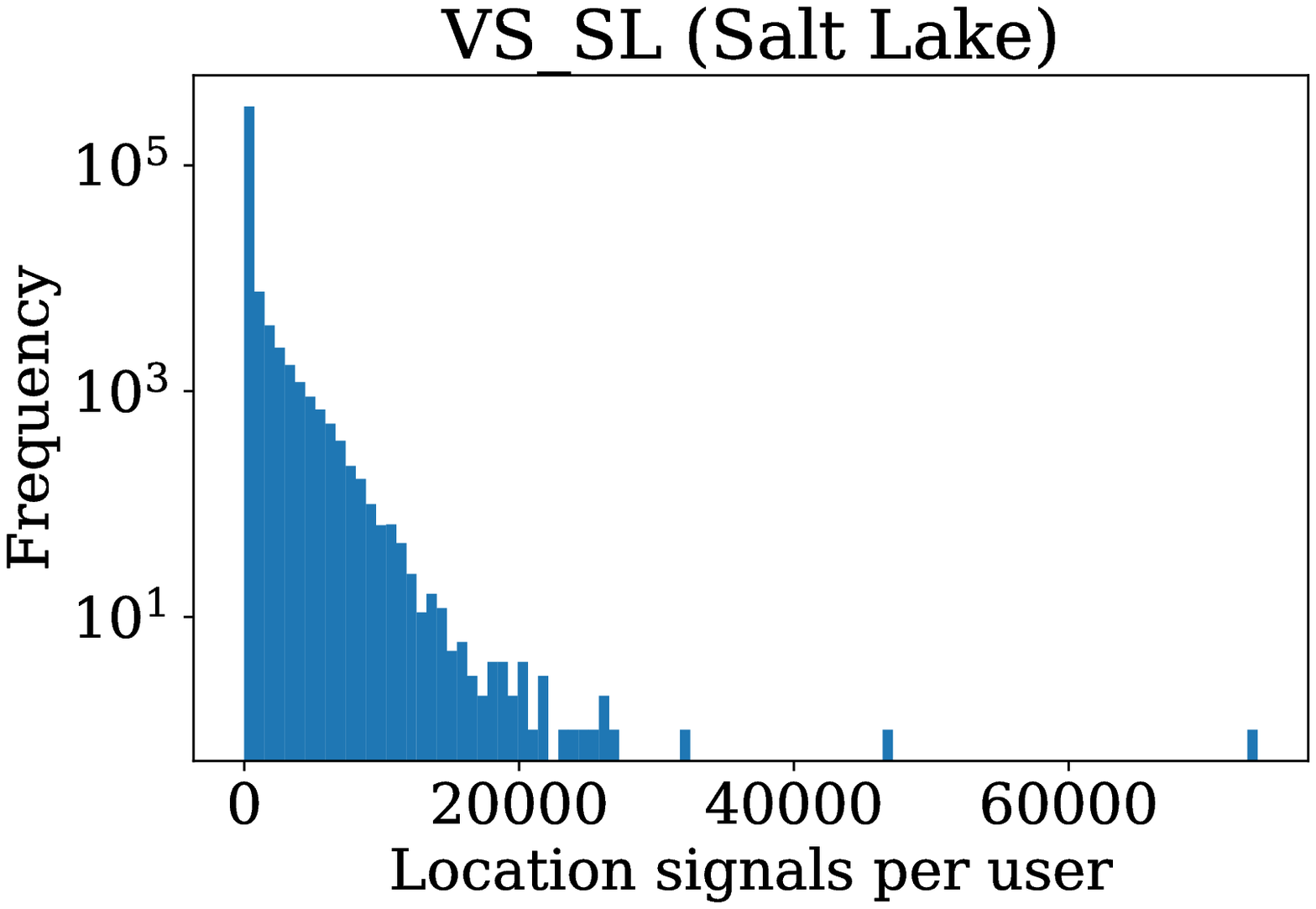}
    \end{minipage}
    \begin{minipage}[t]{0.23\textwidth}
        \includegraphics[width=\textwidth]{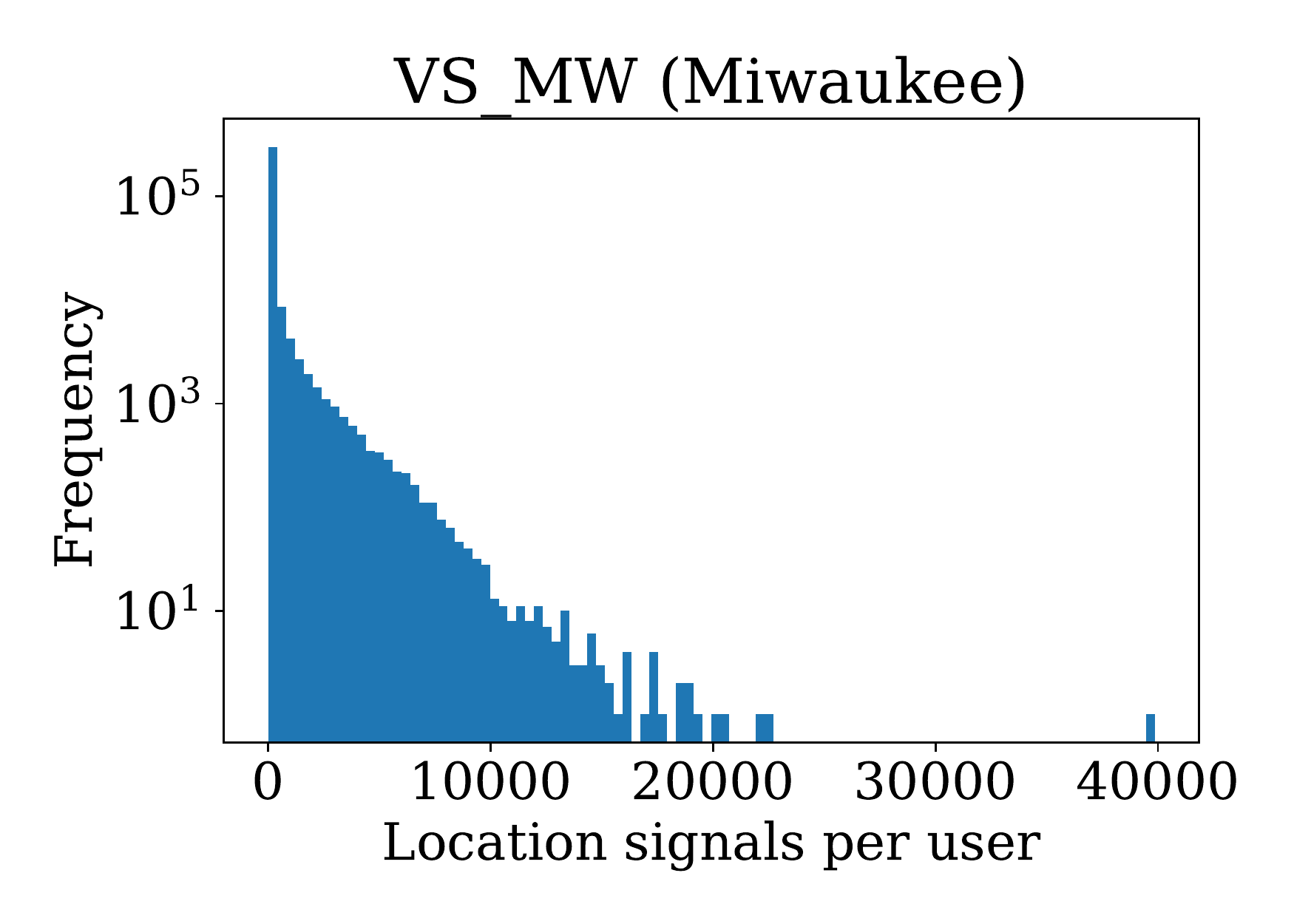}
    \end{minipage} 
    \caption{Histogram of location signals per user in proprietary datasets}
    \label{fig:datasets_power_law_vs}
\end{figure*}

Figure \ref{fig:datasets_power_law} shows the histogram of the points contributed by a user for datasets 4SQ\_TKY, GW\_SF and GW\_NY. Figure \ref{fig:datasets_power_law_vs} shows the histogram of the points contributed by a user for datasets VS\_HT, VS\_LA, VS\_SL and VS\_MW.

We report power-law curve fit coefficients for each of the veraset dataset, as fit to the power-aw function $f(x) = a \times x^b$ in Table \ref{tbl:table_power_law}.


\end{document}